# Isotropic turbulence in compact space

Elias Gravanis[1] and Evangelos Akylas[1]

[1]*Department of Civil Engineering and Geomatics,*
*Cyprus University of Technology*
*P.O. Box 50329, 3603, Limassol, Cyprus*
e-mail: elias.gravanis@cut.ac.cy;  evangelos.akylas@cut.ac.cy

## ABSTRACT

Isotropic turbulence is typically studied numerically through the direct numerical simulations (DNS). The DNS flows are described by the Navier-Stokes equation in a 'box', defined through periodic boundary conditions. Ideal isotropic turbulence lives in infinite space. The DNS flows live in a compact space and they are not isotropic in their large scales. Hence, the investigation of important phenomena of isotropic turbulence, such as anomalous scaling, through the DNS is affected by large scale effects in the currently available Reynolds numbers. In this work we put isotropic turbulence – or better, the associated formal theory – in a 'box', through imposing periodicity at the level of the correlations functions. This is an attempt to offer a framework where one may investigate isotropic theories/models through the data of DNS in a manner as consistent with them as possible. We work at the lowest level of the hierarchy, which involves the two-point correlation functions and the Karman-Howarth equation. Periodicity immediately gives us the discrete wavenumber space of the theory. The wavenumbers start from 1.835, 2.896, 3.923, and progressively approach integer values, in an interesting correspondence with the DNS wavenumber shells.

Unlike the Navier-Stokes equation, infinitely smooth periodicity is obstructed in this theory, a fact expressed by a sequence of relations obeyed by the normal modes of the Karman-Howarth equation at the end-points of a unit period interval. Similar relations are imparted to the two-point functions under the condition that the energy spectrum and energy transfer function are realizable. Hence these relations are necessary conditions for realizability in this theory. Naturally constructed closures scheme for the Karman-Howarth equation do not conform to such relations, thereby destroying realizability. A closure can be made to conform to a finite number of them by adding corrective terms, in a procedure which possesses certain analogies with the renormalization of quantum field theory. Perhaps the most important one is that we can let the spectrum to be unphysical (through sign-changing oscillations of decreasing amplitude) for the infinitely large wavenumbers, as long as we can controllably extend the regime where the spectrum remains physical, deep enough in the dissipation subrange so that to be realistically adequate. Indeed, we show that one or two such 'regularity relations' are needed at most for comparisons of the predictions of the theory with the current resolution level results of the DNS. For the implementation of our arguments we use a simple closure scheme proposed in the past by Oberlack and Peters. The applicability of our ideas to more complex closures is also discussed.

## I. INTRODUCTION

Isotropic turbulence is an ideal type of flow which is characterized by the invariance of the scalar correlations functions under translation and rotation. Isotropic turbulent flows live in infinite space. A bounded domain does not realize isotropy as there are points not belonging to the domain in every direction around a boundary point. Hence, not all directions are equivalent in a bounded domain. The problem is partially solved by introducing periodicity. One works with a cubic domain and insists that the velocity field at opposite faces of the cube is the same. That means that exiting the cube from one face we enter the cube from the opposite face or, equivalently, that we have maintained the infinite space only we have divided in identical cubic cells. This is the space where the direct numerical simulation (DNS) flows live [1]. It is a space where local isotropy is not lost. Periodicity allows us to imitate infinite space in a finite one by removing both formally and intuitively any notion of boundary. Nonetheless the problem remains globally. In the large scale, not all directions of the cube are equivalent. The large scales of the DNS flows are not isotropic. The problem is solved by simply accepting it, and focusing on the smaller scales which are increasingly isotropic the smaller they are.

This practice would be perfectly fine if the available Reynolds numbers in the DNS were really very large. Then the largest wavenumbers phenomena, which are due to the lack of isotropy and exaggerated by the fact that energy is fed in the system precisely there, would not be very important. In the current DNS, important phenomena such as the anomalous scaling are read off from the first few dozen wavenumbers which are impossible to cleanly separate from the effects of the anisotropic large scales and the effects of forcing [2-4]. Nonetheless, theoretical analyses regarding these phenomena must necessarily be done within the usual isotropic turbulence formalism, as it is precisely this kind of theories one would like to test. Now, although we cannot turn the DNS flows to ideal isotropic ones, we could possibly make the classical formalism to at least accommodate consistently the existence of a permanent largest scale, that is, a sort of boundedness of the domain, most conveniently implemented by periodicity. Then the investigation of the DNS flows, through modeling combined with theoretical ideas, can be cast in a language which is more natural to these flows, without losing its sanctified and highly convenient isotropic character.

Once set in this course, one finds that the problem at hand is a fine mathematical problem which is interesting in its own right. A single basic idea implies a number of things, which are part of the final solution to the problem, but also produces a few puzzles. The puzzles are then solved by merely applying consistently the implications and the tools provided by that very basic idea. The idea is to impose periodicity at the level of the correlations functions of isotropic turbulence, being indifferent to what underlying velocity field could produce such functions. Due to the hierarchy problem one restricts oneself to the two-point correlation functions and the Karman-Howarth equation [5][6]. The two-point functions live on an infinite line which periodicity turns the infinite line into a circle, parameterized by $-\pi \leq r \leq \pi$. Also the continuous spectral space becomes discrete. The large wavenumbers approach integers but the small wavenumbers $k$ start from 1.835, 2.896, 3.923 and so on, creating an interesting correspondence with the wavenumber shells of the DNS flows, where the low wavenumber ones are under-populated by wavenumber vectors and their labeling by integers is only nominal. Closed form formulas relating the periodic longitudinal autocorrelation function to the discrete energy spectrum as well as the triple correlation function to the energy transfer, are explicitly derived, through expansions on the modes corresponding to the wavenumbers.

At this point an interesting subtlety arises. The modes, which are naturally built for the infinite space, they allow a periodic function to be constructed out of them which is smooth only up to the second derivative, same as a periodic function built out of parabolic segments. This deviation from the infinitely smooth periodicity is encoded in a sequence of equations which relate all odd order derivatives of the modes at $r=\pi$ to lower order even derivatives at that point; the modes are even functions, thus all their odd derivatives would vanish at that point if they were infinitely smooth e.g. as it is the case of the function $\cos(r)$. These relations imply a sequence of similar relations on the two-point functions when combined with the condition of physically realizable energy spectrum and energy transfer function, understood as positive and decreasing for large wavenumbers so that all their moments exist. That is, the periodic autocorrelation function and the triple point function must obey a series of constraints involving increasingly higher derivatives at $r=\pi$, which we call as the regularity conditions. A spectrum is physical for arbitrarily large wavenumbers, if all these relations are satisfied by the autocorrelation function. The spectrum becomes asymptotically oscillatory when a finite number of them are satisfied; specifically, if $n$ conditions are satisfied (which what we call as order $n$ regularity) then the dominant large $k$ behavior of the spectrum is given by $(-1)^{[k]}/k^{2n+2}$, where $[k]$ is the integer part of $k$. There is a couple of interesting facts about this behavior. First, the unphysical behavior starts deep enough in the dissipation subrange so that the unphysical part of the spectrum is not of practical relevance. Second, in spite of their power law asymptotic behavior, this kind of spectra allows all the energy moments to be definable in a weak sense. The behavior of the energy transfer in its relation to the periodic triple correlation follows a similar pattern.

The Karman-Howarth equation cannot become a closed equation unless we employ a modeled functional relation between the triple correlation function and the autocorrelation function, that is, use a closure scheme. A closure scheme turns the Karman-Howarth equation to a diffusion type equation for the autocorrelation function such that it involves a non-linear self-interacting term (the closure term). A closure scheme is naturally constructed to make sense in infinite space. Therefore, it does not conform to, and does not preserve the regularity of the autocorrelation function while operating as the interacting part of the Karman-Howarth equation. The remedy to this problem is to add corrective terms to the closure so that it conforms to regularity order by order. It turns out that one or two orders of regularity are adequate to give us a physical spectrum as a solution of the Karman-Howarth equation far enough in the dissipation subrange so that model predictions can be compared with DNS data of the current resolution level. Put differently, periodicity introduces another kind of resolution level, i.e. the point where the irregularities of the spectrum start occurring, which, through a systematic procedure, can be pushed to large enough wavenumbers so that to be practically unimportant. This procedure exhibits quite a few analogies with the renormalization theory of quantum field theory; hence, we borrow various terms for our procedure from renormalization including its name. Summarizing, periodicity creates a puzzle through the asymptotically oscillatory spectra it produces, but also provides the solution through the regularity conditions and the effectiveness of their use through the procedure we have called as renormalization.

The development of our construction requires quite a bit of formal work to be presented, hence we shall not dwell also on the development of closure schemes. For illustration purposes we will simply use a decently behaving existing model, developed in the past by Oberlack and Peters [7], built solely on the basis that it relates consistently the Kolmogorov laws for the second and the third structure functions in the inertial subrange. Hence we shall solve the Karman-Howarth equation for this closure, using existing DNS input data, as an exercise that illustrates how this machinery we have built operates in this problem and, primarily, how our renormalization procedure works, studying in adequate detail the associated phenomena. On general physical grounds, one should bear in mind that a closure scheme is a low order approximation of the infinite hierarchy of the correlation functions dynamical equations, and as such is not complete in any sense. Therefore, it is not at all necessary for a closure to be perfect from a specific point, for example, to produce a physical spectrum for infinitely high wavenumbers, as long as it gives meaningful answers to meaningfully stated questions. Such questions involve comparison of the theoretical results to actual data where a given resolution scale is always at work.

In section II we give some basic formulas of isotropic turbulence. In section III we discuss the passage from the Navier-Stokes in a box (DNS flows) to the Karman-Howarth in box which is our new doctrine, through the common theme of imposing periodicity, or 'compactifying' the corresponding theories, as we put it. In section IV we develop the details of the theory. In section V we discuss certain aspects of the correspondence between the present theory and the DNS flows. In section VI we apply the results in a situation that motivated this work, (forced) turbulence in the DNS, showing how that can be transcribed in our context and results can be derived. Emphasis is given on the effects of renormalization rather than on phenomenology. In section VII we give an overview of the theory from the point of view of the finished product, and also discuss the application of the presented ideas in more complex closures required by phenomenology.

## II. ISOTROPIC TURBULENCE

We start by reviewing certain facts about isotropic turbulent flows we shall need in what follows. Incompressible turbulent flows are fundamentally described by the Navier-Stokes equation along with the incompressibility constraint:

$$\frac{\partial u_i}{\partial t} + u_k \frac{\partial u_i}{\partial x_k} = -\frac{1}{\rho}\frac{\partial p}{\partial x_i} + \nu \frac{\partial^2 u_i}{\partial x_k \partial x_k}, \qquad \frac{\partial u_i}{\partial x_i} = 0 \qquad (1)$$

where $u_i$ is the Eulerian velocity field and $p$ the pressure. Due to the complexity of the system the physically relevant information about the flow is of statistical nature. Denoting a suitable statistical ensemble average by angle brackets, the tensor field $\langle u_{i_1}(\vec{x}_1) u_{i_2}(\vec{x}_2) \cdots u_{i_n}(\vec{x}_n) \rangle$ is the same-time, order $n$ correlation function. Some of the points $x$ may coincide, so that there are only $m<n$ different $x$'s involved and these fields may be designated as $m$-point, order $n$ correlation functions.

We consider flows with zero mean velocity, $\langle u_i(\vec{x})\rangle = 0$. The simplest non trivial single-point correlation functions of the velocity field are the total kinetic energy $K = \frac{1}{2}\langle u_i u_i\rangle$ and the dissipation $\varepsilon = \nu\langle \partial_k u_i \partial_k u_i\rangle$. Consider ideal homogeneous turbulent flows. Then the single-point correlation functions do not depend on the position in space. Therefore correlation functions which are total derivatives vanish identically. This allows us to show that $K$ and $\varepsilon$ are related by the exact energy balance equation

$$\frac{dK}{dt} = -\varepsilon \qquad (2)$$

Consider ideal isotropic turbulent flows. There is no intrinsic direction in the flow, locally or globally. These flows are automatically homogeneous. Ideally homogeneous and therefore isotropic flows exist necessarily in an infinite space. The correlation function tensor fields can be reduced purely geometrically to a set of scalar fields that depend only on rotation-invariant quantities, distances and angles. In the place of scalar fields one usually uses pseudo-scalar fields with specific parity properties. For simplicity we may refer also to them as 'scalar fields'.

Any $m$-point correlation function $\langle u_{i_1}(\vec{x}_0) u_{i_2}(\vec{x}_1)\cdots\rangle$ depends on the separations $\vec{r}_a = \vec{x}_a - \vec{x}_0$, where $a = 1,\ldots,m-1$. This function can be reduced to the associated longitudinal correlation scalar fields

$$(\vec{e}_a)_{i_1}(\vec{e}_b)_{i_2}\cdots\langle u_{i_1}(\vec{x}_0) u_{i_2}(\vec{x}_1)\cdots\rangle \qquad (3)$$

where $\vec{e}_a$ is a unit vector parallel to $\vec{r}_a$. Transverse fields are obtained by replacing some of the $e$'s with vectors perpendicular to them. The scalar fields depend only on the lengths of the $\vec{r}_a$ and the angles between them. It is not hard to see that the natural domain of an $m$-point correlation function is an $(m-1)$-dimensional hyperplane.

The correlation functions obey a set of infinite, strongly coupled equations of motion. The equations of motion of the $m$-point, order $n$ function involve the $(m+1)$-point and order $(n+1)$ function restricted on the $(m-1)$-dimensional domain of the order $n$ function. This applies for all $n \geq 2$. This is the hierarchy of correlation function equations. The coupling follows from the quadratic term of the Navier-Stokes equation. This gives rise to the 'closure problem'. Closing off the infinite sequence of equations at some finite order has to be done semi-empirically.

An exceedingly simple choice is to adopt a closure scheme that allows us to work solely at the lowest non-trivial level of the hierarchy i.e. the equations of motion of the order-two function. Quite a few interesting things can be learned from that low level of the hierarchy, entirely independently of the closure scheme, that can possibly be used at the higher levels.

The order-two correlation function can be reduced, by isotropy and incompressibility, to the longitudinal scalar field $F(r) = \langle u_l(0) u_l(\vec{r})\rangle$ where $u_l$ is the component of the velocity field along an axis that is parallel to $\vec{r}$ and has a fixed positive sense. The field $F(r)$ is an even function over the line $-\infty < r < \infty$.

By the velocity field equations (1) one derives the *Karman-Howarth equation* [5]

$$\frac{\partial F}{\partial t} = \frac{1}{r^4}\frac{\partial}{\partial r}\left[r^4\left(2\nu\frac{\partial F}{\partial r} + H\right)\right] \tag{4}$$

It is the dynamical equation for $F$, a linear diffusion type of equation with an interaction term involving the field $H$. The field $H$ couples the Karman-Howarth equation, and the field $F$, to the rest of the hierarchy. Explicitly, $H(r) = \langle u_l(0)u_l(0)u_l(\vec{r})\rangle$ and it is an odd function. The field $H$ is the restriction on the line $-\infty < r < \infty$ of a scalar field that lives on a plane.

The point $r = 0$ is a singular point of equation (4). That is, auxiliary conditions specifying the local behavior of the fields in the neighborhood of $r = 0$ are required. The smoothness of the velocity field obeying the Navier-Stokes equation implies that the correlation functions are also smooth. Therefore evaluating a quantity at $r = 0$ is the same as calculating its limit for $r \to 0$. Noting also that incompressibility requires that $\partial_r H |_{r=0} = 0$ the limit of equation (3) for $r \to 0$ implies that

$$\left.\frac{\partial F}{\partial t}\right|_{r=0} = -10\nu\left.\frac{\partial^2 F}{\partial r^2}\right|_{r=0} \tag{5}$$

This equation is also equivalent to the energy balance equation (2) upon using the formula for the dissipation rate as it is simplified by isotropy

$$\varepsilon = -15\nu\left.\frac{\partial^2 F}{\partial r^2}\right|_{r=0} \tag{6}$$

and observing the elementary fact

$$F(0) = u'^2 = \tfrac{2}{3}K \tag{7}$$

where $u'^2 = \langle u_l(0)u_l(0)\rangle = F(0)$ is the one-directional mean squared of the velocity field.

**III. COMPACTIFICATION AND TURBULENCE IN THE DNS**

The isotropic flow, requiring first of all an infinite physical space to live in, is an idealization that cannot be strictly realized even in the clinical environment of numerical simulations. In the direct numerical simulations (DNS) of isotropic turbulence one usually solves the Navier-Stokes equation imposing periodicity. The idea is to introduce finiteness in space in a smooth manner. Isotropic flows are replaced by another kind of ideal flows that can be handled numerically. Explicitly, the DNS turbulent flows are governed by the velocity field equations (1) satisfying the conditions

$$u_i(x, y, z) = u_i(x+l, y, z) = u_i(x, y+l, z) = u_i(x, y, z+l) \qquad (8)$$

for the axes $x$, $y$, $z$ where we explicitly write down (1) in components.

Condition (8) introduces a tiling, or tessellation, of the infinite physical space. The tiling is made up of an infinite number of cubic domains of side $l$. Each such cubic domain is a unit cell of the tessellation repeated infinitely many times. In practice, one usually thinks in terms of a single such unit cell speaking of 'periodic boundary conditions'. This is slightly misleading. Each unit cell of the tessellation is a notional subset of the infinite space that does not bound the flow in its interior. The condition (8) does not change if we pick another origin for the axis $x$, $y$ and $z$ as long as we do not rotated those axes. The condition (8) does not distinguish between tessellations whose unit cells are parallel. The boundary of the unit cell of a given tessellation contains points that lie in the interior of a unit cell of another equally good tessellation.

We may adopt a more abstract view. The condition (8) says that the field configuration looks the same at all points in space whose coordinates differ by integer multiples of $l$. This is why we can restrict ourselves into a unit cell. We lose no bit of information about the field if we identify all points in space whose coordinates differ by integer multiples of $l$. That means that all unit cells are mapped on a single one whose boundary points have been identified. The result is a compact space without boundary. The condition (8) is reduced merely to the single-valued-ness of the field on the compact space. Clearly this treatment can be applied to any cubically tessellated space. Compactification is the term we shall use for this procedure, implying both the operation on the space itself as well as on the field theory that lives in it. If a field theory lives in an infinite line then the unit cell is an interval whose end-points are identified. The result is a field theory that lives in a circle. If a theory lives on a plane then the unit cell is a square whose opposite sides are identified resulting in a theory that lives in a compact space which topologically is a torus. DNS flows are described by the compactification of the Navier-Stokes (1) in a space which topologically is a three-torus.

The more abstract view just described lies in the difference between the local and the global properties of a space. The tori are intrinsically flat spaces. Locally they are not different than the original infinite spaces. Globally, they are clearly non-trivial. The field theories of interest are written in terms of local equations, such as the Navier-Stokes equation. Therefore compactification does not affect the form of the equations but rather determines the boundary conditions and other global specifications.

DNS turbulence can be regarded as statistically homogeneous turbulence. From the point of view of compactification it is easy to see that there is no a priori difficulty with homogeneity: The compact space is perfectly homogeneous and the flow encounters no special points. Regarding isotropy things are different. The compact space does have special directions; to assist imagination one may use the two-torus as a model of that space. The space is perfectly isotropic locally, allowing the flow to evolve and adjust itself accordingly. Globally it is not, affecting the overall state of isotropy in the flow as scales are coupled to each other through cascade. One may note that the solely global breaking of isotropy is no surprise since compactification is a strictly global operation on a perfectly homogeneous and isotropic space. In all, DNS turbulent flows are certainly not ideally isotropic but due to the compromise between the local and global properties of the compact space these flows acquire a considerable degree of isotropy.

Now the DNS flows are supposed to simulate ideal isotropic flows in the first place, and we want to use the formulas valid in the latter with all their irreplaceable simplicity, and have at our disposal the suitable framework for studying fundamental conjectures such as Kolmogorov's theory or deviations from it. In practice, one proceeds by eliminating the direction-dependence of the numerical results by averaging over the directions; specifically this is done in the Fourier space in quantities such as the energy spectrum. The resulting scalar quantities are then regarded as adequate approximations of the analogous quantities in ideal isotropic flows for scales adequately smaller than $l$. Indeed, in practice, standard measures of isotropy show small, fluctuating deviations around their ideal values. The whole thing suggests then a rather interesting theoretical problem. Can we set up a framework that maintains all the elegance and simplicity of the ideal isotropic flows - at the formal level - that also reflects compactness of space which is the defining property of the DNS flows?

## IV. ISOTROPIC TURBULENCE IN COMPACT SPACE

In few words, instead of compactifying the domain of the velocity field we compactify the natural domains of the correlation scalar fields of isotropic flows. That is, locally in their natural domains the correlation scalar fields are governed by all the usual equations but their global specifications are modified.

### A. Compactification and boundary conditions

For every correlation scalar field of isotropic turbulence we introduce an abstract scalar field satisfying all formal relations satisfied by the correlation scalar field locally in its natural domain. In particular, we introduce an abstract field $F$ which is an even function of a coordinate $r$ on an infinite line and an abstract field $H$ which is an odd function on that line, satisfying all relations from (2) to (7). Predominantly, the Karman-Howarth equation (4) is satisfied. The relation (7) is regarded essentially as a definition of the kinetic energy $K$, and (2) or (6) are equivalent definitions of the energy dissipation rate $\varepsilon$.

A priori the natural domains of the abstract scalar fields are lines, planes and hyperplanes, subspaces of an infinite dimensional hyperplane. Then we compactify, in a way that preserves flatness: We impose that infinite lines are topologically 'wound' to circles, planes to two-tori and hyperplanes to $n$-tori, subspaces of an infinite dimensional torus. Explicitly that means imposing periodicity on all abstract scalar fields, a multi-dimensional analog of the periodicity condition (8). As discussed in section II, our mathematical set up inherits the so-called closure problem, meaning that the interactions are non-trivially entangled, and we probably need the full set of the infinite scalar fields in order to answer fundamental questions from first principles; hence, we restrict ourselves to the lowest order, i.e. the one-dimensional components of the full theory and study its compactification to a circle.

The infinite line $-\infty < r < \infty$ is compactified to a circle of circumference $2\pi$. We shall work in the interval $-\pi \leq r \leq \pi$. The circumference or period $l=2\pi$ simply reflects the usual choice in the DNS flows; it can be changed to any value by rescaling $r$. The dynamical equation in the problem, the Karman-Howarth equation, is second order in the field $F$ and first order in the field $H$. Therefore, the boundary conditions encoding periodicity on the fields $F$ and $H$ in this interval are

$$F(-\pi) = F(\pi), \qquad \partial_r F(-\pi) = \partial_r F(\pi)$$
$$H(-\pi) = H(\pi) \tag{9a}$$

The field $F$ is even function, which means that $\partial_r F$ is odd, and the field $H$ is odd. Thus the first condition on $F$ is trivially satisfied and the remaining two conditions in (9a) read equivalently

$$\partial_r F(\pi) = 0, \qquad H(\pi) = 0 \tag{9b}$$

A most fundamental fact which will be realized in the course of our discussion, is that although the fields $F$ and $H$ must be infinitely differentiable everywhere in $-\pi \leq r \leq \pi$, they are only twice and once differentiable respectively, if one indeed views this interval as a parameterization of the circle where the point $r=-\pi$ is identified with the point $r=+\pi$: Higher derivatives of the fields become discontinuous as one crosses the point $-\pi \sim +\pi$. Employing the notion of $C^k$ differentiability class (that is, $k$ times continuously differentiable) we may say that the field $F$ is $C^2$ class periodic while $H$ is $C^1$ class periodic. Such functions are not unusual: they occur every time one creates a periodic function by piecing together copies of the same segment of the curve of e.g., a polynomial, because the matching is only finitely smooth at the joining points, that is, as one joins one boundary point of the curve segment to the other. In our case the 'failure' of higher order periodicity is encoded in a sequence of relations on the fields $F$ and $H$ which are fundamental for the completion of our theory. These relations follow once one develops the spectral version of the theory, to which we now turn.

**B. The spectral space and the dynamical degrees of freedom**

The abstract field $F(r)$ should give rise to an abstract energy spectrum $E(k)$. Constructing it with all the associated details is the main subject of this subsection. Of course we need first to find out what $k$ is.

The spectral space is associated with the normal modes of the theory of $F$ in the compact space, determined by the linear in $F$ i.e. the non-interacting part of the Karman-Howarth equation in that space. This is a neat answer that may sound both perfectly natural or ad-hoc depending on whether one doubts the relevance to the problem of the splitting of the field equation into linear (free) and interacting part. For that reason we prefer to show that it is indeed the correct answer.

We may put our thoughts in order by returning to the ideal isotropic flows. The relation between the autocorrelation function $F(r)$ and the energy spectrum $E(k)$ in isotropic turbulence is

$$F(r) = \tfrac{2}{3}\int_0^\infty dk\, E(k)\, \frac{3}{(kr)^2}\left(\frac{\sin kr}{kr} - \cos kr\right) \tag{10}$$

$k$ is the magnitude of the wave-number vector $k_i$. The domain of the ideal flows is the infinite space therefore the wave-number space is continuous and infinite itself.

Using Fourier transforms in the tensorial quantities in three-dimensions means that the normal modes are the usual plane waves $\exp(i\vec{k}\cdot\vec{r})$, leaving no mystery to what spectral space means. On the other hand, at the level of scalar quantities that solely interest us, one is left with (10). Is expression (10) itself an expansion of the field $F$ on normal modes? It may be so if the functions

$$f_k(r) = \frac{3}{(kr)^2}\left(\frac{\sin kr}{kr} - \cos kr\right) \tag{11}$$

are the normal modes of some fundamental dynamical equation in the problem. Note that that $f_k(r)$ are normalized so that $f_k(r) \to 1$ for $r \to 0$.

Indeed, $f_k(r)$ are eigenfunctions of the linear operator $r^{-4}\partial_r\, r^4 \partial_r$ which acts on the field $F$ in the Karman-Howarth equation (4), which we may re-write as

$$\frac{\partial F}{\partial t} = 2\nu\, \frac{1}{r^4}\frac{\partial}{\partial r}\, r^4\, \frac{\partial}{\partial r} F + \text{interacting part}$$

The eigenvalue equation for this linear operator is

$$\frac{1}{r^4}\frac{\partial}{\partial r}\, r^4\, \frac{\partial}{\partial r} f_k(r) = -k^2 f_k(r) \tag{12}$$

The eigenvalues $-k^2$ fix the time-dependence of the respective modes $f_k(r)$, $\exp(-2\nu k^2 t)$. As anticipated above, the normal modes are indeed specified most naturally by the free part of the field equations.

Equation (12) is a Sturm-Liouville eigenvalue equation [8], and $f_k(r)$ are its solutions which are analytic around the origin. No special conditions need to be imposed at the 'boundary' $r = \pm\infty$ thus there are no restrictions on the values of $k$. The spectral space is continuous. Equation (10) is an expansion of $F(r)$ on the modes $f_k(r)$ for all $k$. The coefficients form the energy spectrum, $E(k)$. Equation (11) is consistent with (7) by $\int_0^\infty dk\, E(k) = K$.

Compactifying the infinite line to a circle of circumference $2\pi$ means that the modes $f_k(r)$ satisfy the eigenvalue equation (12) everywhere in $-\pi \le r \le \pi$ and satisfy the boundary conditions on $F$ given by (9a) or equivalently (9b):

$$f_k'(\pi) = 0 \tag{13}$$

bearing in mind that the modes (11) are even functions same as the field $F$. We should explicitly include the zero-mode $f_0(r)=1$ whose existence is downplayed in the infinite line case. All modes are square integrable in the interval $-\pi \le r \le \pi$. The conditions (13) specify the eigenvalues $-k^2$, in other words, they specify the spectral space of the compactified theory.

The spectral space is discrete, a result of the compactness of the domain. Explicitly, equation (13) gives that $k$ is given by the solutions of the transcendental equation

$$(k^2\pi^2 - 3)\sin k\pi + 3k\pi \cos k\pi = 0 \tag{14}$$

The first few solutions of (14) are $k = 0$ and

$$k = 1.835,\quad 2.896,\quad 3.923,\quad 4.939,\quad 5.949 \tag{15}$$

One may fact show (see Appendix A) that the $n$-th non-zero wavenumber $k$ is the $n$-th zero of the Bessel function of the first kind of order 5/2: $J_{5/2}(\pi k)=0$.

One observes that there is no value in the vicinity of $k = 1$. Also there is that $k = 0$ mode. Both phenomena admit interesting interpretations that we will discuss below. One also observes that values of $k$ are not integers but they are in the vicinity of integers and approach them asymptotically. The large values of $k$ are given by

$$k = n - \frac{3}{\pi^2 n} + O(\frac{1}{n^2}) \tag{16}$$

for large integers $n$. The properties of $k$ are discussed in more detail in the Appendix A.

We have therefore determined completely the normal modes of the compactified theory. We may finally write down the mode expansion of the field $F$ in the compact space:

$$F(r) = \tfrac{2}{3}\sum_k E(k) f_k(r) \tag{17}$$

where the sum is taken over the non-negative solutions $k$ of (14). $E(k)$ is defined by this expansion as the abstract energy spectrum. By the relations (7) and (6) we obtain

$$K = \sum_k E(k), \qquad \varepsilon = 2\nu \sum_k k^2 E(k) \tag{18}$$

These relations have the form expected by the interpretation of the quantities involved.

The modes $f_k(r)$ are eigenfunctions of a Sturm-Liouville problem and therefore are orthogonal; orthogonality of course holds in the $\delta$-function sense in the infinite line case as well. We make orthogonality explicit in order to invert the expansion (17). For any positive $k$ and $k'$ we find (Appendix B)

$$\int_{-\pi}^{\pi} f_k(r) f_{k'}(r) r^4 dr = \delta_{kk'} \frac{9\pi}{k^4} c(k), \qquad c(k) = 1 - \left\{\frac{1}{3} + \frac{1}{(k\pi)^2}\right\} \sin^2 k\pi \tag{19}$$

For the zero-mode $f_0(r) = 1$ and any positive $k$ mode $f_k(r)$ orthogonality reads

$$0 = \int_{-\pi}^{\pi} r^4 f_k(r) f_0(r) dr = \int_{-\pi}^{\pi} r^4 f_k(r) dr \tag{20}$$

These equations are an immediate implication of the eigenvalue equation (12) and the boundary conditions (13).

We are now able to deduce an explicit expression for the abstract energy spectrum as a transform of $F$:

$$E(k) = \frac{1}{2\pi c(k)} \int_{-\pi}^{\pi} \left(\frac{\sin kr}{kr} - \cos kr\right) (kr)^2 F(r) \tag{21}$$

for any positive $k$, and

$$E(0) = \frac{15}{4\pi^5} \int_{-\pi}^{\pi} r^4 F(r) dr \tag{22}$$

for $k = 0$. We learn most convincingly that the zero-mode not only cannot be lightheartedly discarded within the discrete spectrum of the compactified theory but also its energy is proportional to what one would reasonably call as the Loitsyansky integral in this context. We also learn that the zero-mode is the sole contributor to the Loitsyansky integral, as it is most clearly expressed by (20). We will demystify these highly important properties of the compactified theory below. We shall also drop the adjective 'abstract' from now on for the fields $F$, $H$ and $E(k)$, understanding that we refer to the compact space analogues of the autocorrelation function, triple correlation function and the energy spectrum.

Inserting the mode expansion (17) of the field $F$ into the Karman-Howarth equation (4) we obtain a set of dynamical equations for the spectrum $E(k)$. At this point the fact that $f_k(r)$ are the normal modes of the free (non-interacting) part of the Karman-Howarth comes into play and 'diagonalizes' the free part of the equations for $E(k)$. For all positive $k$ modes one finds

$$\frac{dE(k)}{dt} = -2\nu k^2 E(k) + T(k) \tag{23}$$

where the energy transfer $T(k)$ is given by

$$T(k) = -\frac{1}{6\pi c(k)} \int_{-\pi}^{\pi} H(r)(kr)^4 f'_k(r) dr \tag{24}$$

The orthogonality relation for $f'_k(r)$ is similar to (19) times a factor $k^2$ (relation (B5)). This allows us to invert (24) to write the field $H(r)$ as an expansion of the modes $f'_k(r)$:

$$H(r) = -\frac{2}{3} \sum_k \frac{1}{k^2} T(k) f'_k(r) \tag{25}$$

$H(r)$ is an odd function, same as the modes $f'_k(r)$. By the condition (13) the field $H$ automatically satisfies the boundary condition (9b). The conservation of energy transferred between the modes, expressed by $\sum_k T(k) = 0$, is equivalent to $H'(0)=0$. The latter condition is a well-known fact regarding the triple correlation function in isotropic turbulence [6] which follows explicitly from the incompressibility condition on the velocity field. The sum in (25) excludes of the zero-mode to which we now turn.

The energy $E(k)$ of all positive $k$ modes decays through dissipation at a rate $2\nu k^2 E(k)$ and it is transferred to other modes through $T(k)$ that expresses the cascade. For the zero-mode we have

$$\frac{dE(0)}{dt} = 0 \tag{26}$$

The zero-mode energy does not decay through dissipation, as the associated rate $2\nu k^2 E(k)$ vanishes, and it is not transferred to other modes either, unless we can allow for a non-zero value for $T(0)$. We will be explicitly prove (26) by a different route below.

With the aid of the last result we may re-write the mode expansion (17) as follows:

$$F = \text{constant} + \tfrac{2}{3} \sum_{k>0} E(k) f_k(r), \qquad \text{constant} = \tfrac{2}{3} E(0) \tag{27}$$

Therefore the 'vacuum' configurations, $F$=constant and $H$=0, are associated with the zero-mode. Indeed, if we proceed inversely and put $F$=constant into the formula (21) the orthogonality relation (20) tells us that $E(k)$=0 for all $k>0$.

The transformation

$$F \to F + c_1, \qquad H \to H \tag{28}$$

where $c_1$ is a constant, leaves the Karman-Howarth equation and the boundary conditions unchanged. That is, this transformation is a symmetry of the theory mapping a solution of the equations to another solution. The entirety of the 'vacua' $F$=constant and $H$=0 are generated from the solution $F$=0 and $H$=0 by the transformation (28). What is more, the transformation (28) acts on the general configuration (26) by simply moving the 'vacuum':

$$\text{constant} \to \text{constant} + c_1, \qquad E(k) \to E(k) \tag{29}$$

The integral in the right hand side of (21) cannot 'see' additive constants in $F$. In more words, varying the zero-mode energy $E(0)$ of an arbitrary configuration we obtain another configuration that differs *only by an additive constant* in the field $F$. We conclude then that the zero-mode is not only a 'vacuum' mode, but also is completely decoupled from the other modes of the system: $E(0)$ cannot affect the evolution of the system in any way and sets only a bottom for the total energy $K$. One loses nothing by setting that bottom to zero, $E(0)$=0.

What is a little funny is that $E(0)$ is also a proper integral of motion. Formula (22) reminds us that $E(0)$ is essentially the Loitsyansky integral of the theory; the proportionality constant involves the fixed circumference $l$ chosen here to be equal to $2\pi$. Therefore the Loitsyansky integral also is constant throughout motion. This result clearly expresses a conservation law of the theory that it is derivable directly at the level of the field equations for the given boundary conditions. Indeed multiplying the Karman-Howarth equation (4) with $r^4$ and integrating we have that

$$\frac{d}{dt}\left\{\int_{-\pi}^{\pi} r^4 F(r) dr\right\} = \pi^4 \left[2\nu \, \partial_r F(\pi) + H(\pi)\right] - \pi^4 \left[2\nu \, \partial_r F(-\pi) + H(-\pi)\right] \tag{30}$$

and by the periodicity conditions (9) the right hand side vanishes identically, proving explicitly equation (26).

Now our previous conclusions that the specific value of $E(0)$ is of no particular consequence and we may choose to work with fields $F$ such that $E(0) = 0$, suggest that the specific value of the conserved Loitsyansky integral is of no particular consequence either and we may choose initial conditions such that

$$\int_{-\pi}^{\pi} r^4 F(r) dr = 0 \tag{31}$$

The dynamical and interacting positive $k$ modes preserve (31) through time as they have the necessary property built in them, expressed by (20). Generic initial conditions can be generated through the transformation (28) and give rise to solutions that differ from the solutions (31) by an ignorable additive constant in the field F (which equals 2/3 times the energy of the decoupled zero-mode).

### C. Loitsyansky integral and scaling

The conclusion that (31) identifies the entire class of dynamically distinct solutions - modulo irrelevant additive constants - calls for some further discussion. It is well known that the Loitsyansky integral $\int_0^\infty r^4 F(r) dr$ is a quantity with a special significance in the phenomenology of isotropic turbulence, see e.g. [6]. It would be at least instructive if we paused momentarily the formal development of the theory in order to contrast the status of the Loitsyansky integral between the compactified theory and the ideal isotropic flows.

In ideal isotropic flows the physical space is infinite and the spectral space is continuous all the way to $k=0$. The existence of the Loitsyansky integral depends on the behavior of the spectrum $E(k)$ in the infinitesimal neighborhood of $k=0$. When it exists its (possibly approximate) conservation has quite strong dynamical implications. By neat symmetry arguments involving the freedom of rescaling lengths and times at large Reynolds numbers, see e.g. [9] and references therein, one may deduce that the Loitsyansky 'invariant', with dimensions $[L]^7[T]^{-2}$, shapes the evolution of the system such that, for example, the integral scale $L$ evolves by the power law $L \sim t^{2/7}$ at times adequately distant from the initial conditions. Power laws for other characteristic quantities of turbulence, such as the total energy or the Reynolds number, are then also specified.

In the compactified theory there could not be a problem of existence for the Loitsyansky integral and it is exactly conserved. Also there is no such thing as an infinitesimal neighborhood of $k = 0$ and hence the integral is solely determined by the value of energy at $k=0$. The scaling symmetry considerations apply of course in the compactified theory as well but the result is different. In this case there is a constant length in the problem, the fixed circumference $l$ of the compact space. This breaks the symmetry of rescaling lengths that held in the infinite space, and implies that any large length scale $L$ should eventually approach a constant of order $l$ - always for large enough Reynolds numbers.

On the other hand consider times such that $L$ is not as large as $l$ so that re-scaling lengths is still a good symmetry and the scaling arguments we mentioned above apply. Observe also that the very presence of the fixed length l allows one can multiply the Loitsyansky integral with an arbitrary power of $l$ and get an infinity of conserved quantities of dimension $[L]^{\text{any number}}[T]^{-2}$, one of them being the energy $E(0)$ given by (22). In other words we seem to have a wealth of invariants available for scaling arguments. Note also that these additional invariants have come for free. The reason why they came for free is that they mean nothing scale-wise: the Loitsyansky integral is zero, as far as dynamics is concerned. Scaling symmetry arguments applied to any of those invariants produce an empty statement, $L^{\text{any number}} t^{-2} \times 0 = 0$. What predominantly shapes evolution scale-wise in the long run is the compactness of the space itself. On the other hand, other invariants may very well arise during the intermediate stages of the evolution. Nonetheless it seems hard to single out any non-trivial potential invariant in compact space as one can do with the Loitsyansky or the Saffman invariants [6] in the ideal isotropic flows.

**D. Spectrum regularity conditions**

A physically realizable spectrum $E(k)$ of developed turbulence is positive and falls off exponentially for large wave numbers. We shall call such a spectrum as regular or physical; when this notion is applied for finite wavenumbers it will mean that the spectrum is positive and decreasing for large wavenumbers. The compactness of the domain where the field $F$ lives has non-trivial implications when combined with the condition that the associated energy spectrum must be regular.

We start by observing that all the energy moments exist, that is, the quantities

$$\sum_{k} k^{2i} E(k) \tag{32}$$

are finite for all $i=1,2,3,\ldots$. The moments of the spectrum are related to the derivatives of the field $F$ at $r=0$. Indeed, the eigenvalue equation (12) tells us that if we apply $i$ times the operator $-r^{-4}\partial_r r^4 \partial_r$ on an eigenfunction we obtain

$$(-r^{-4}\partial_r r^4 \partial_r)^i f_k(r) = k^{2i} f_k(r) \tag{33}$$

Then, multiplying both sides by $E(k)$, summing over all $k$ and recalling the mode expansion of the field $F$ we find

$$\tfrac{3}{2}(-r^{-4}\partial_r r^4 \partial_r)^i F(r) = \sum_{k} k^{2i} E(k) f_k(r) \tag{34}$$

Taking the limit $r \to 0$ of both sides we have that the right hand side goes to the $i$-th moment given in (32) by the normalization of the eigenfunctions, $f_k(r) \to 1$ for $r \to 0$. The left hand side involves derivatives of the field $F$ up to order $2i$ at $r=0$. The general relation between derivatives of $F$ and the energy moments given by the limit of (34) is a useful formal fact to bear in mind.

In order to make this relation more explicit, we expand the eigenfunctions $f_k(r)$ in power series around $r=0$, to find the relations

$$f_k''(0) = -\tfrac{1}{5}k^2, \quad f_k^{(4)}(0) = \tfrac{3}{35}k^4, \quad f_k^{(6)}(0) = -\tfrac{1}{21}k^6, \ldots \tag{35}$$

The mode expansion (17) of the field $F$ implies then that

$$F''(0) = -\tfrac{2}{15}\sum_k k^2 E(k), \quad F^{(4)}(0) = \tfrac{2}{35}\sum_k k^4 E(k), \quad F^{(6)}(0) = -\tfrac{2}{63}\sum_k k^6 E(k),\ldots \qquad (36)$$

The overall conclusion is that the differentiability of $F$, understood through its mode expansion, rests on the regularity of the spectrum: the left hand side of the relations (34) or (36) makes sense as much as right hand side makes sense. The last statement will be revisited and refined in an interesting way below (subsection E4), dictated by the very properties of the energy spectrum in compact space.

Consider now the 'boundary' point $r=\pi$. Differentiate the equation (33) once and evaluate the result at that point:

$$\left.\partial_r(r^{-4}\partial_r r^4\partial_r)^i f_k(r)\right|_{r=\pi} = (-1)^i k^{2i} f_k'(\pi) = 0 \qquad (37)$$

where the right hand side vanishes by the boundary condition (13). We therefore obtain a sequence of relations for $i=1,2,\ldots,\infty$ the first few of which can be written as

$$f_k^{(3)}(\pi) = -\frac{4}{\pi} f_k''(\pi) \qquad (38a)$$

$$f_k^{(5)}(\pi) = -\frac{8}{\pi} f_k^{(4)}(\pi) + \frac{24}{\pi^3} f_k''(\pi) \qquad (38b)$$

$$f_k^{(7)}(\pi) = -\frac{12}{\pi} f_k^{(6)}(\pi) + \frac{168}{\pi^3} f_k^{(4)}(\pi) \qquad (38c)$$

Either by (37) or by the explicit relations (38) one observes that the values of the odd derivatives of the modes at $\pi$ are related to the values of lower even derivatives at $\pi$ through equations with constant ($k$ independent) coefficients. Explicit relations such as (38) requires using the lower derivatives relations to eliminate all the odd derivative terms which arise in (37) for a given $i$, but we shall not try to give a general simplified result here; the relations (38) will be more than adequate and it is enough to know that general relation is given by (37).

An alternative and somewhat instructive derivation of (38) comes from differentiating explicitly the eigenfunctions as given by equation (11). Indeed, the first few derivatives of the eigenfunctions at $r=\pi$ read

$$f_k(\pi) = \frac{\sin k\pi}{k\pi} \qquad (39a)$$

$$f_k''(\pi) = -k^2 \frac{\sin k\pi}{k\pi} \tag{39b}$$

$$f_k^{(3)}(\pi) = \frac{4}{\pi} k^2 \frac{\sin k\pi}{k\pi} \tag{39c}$$

$$f_k^{(4)}(\pi) = \left\{ k^4 - \frac{24}{\pi^2} k^2 \right\} \frac{\sin k\pi}{k\pi} \tag{39d}$$

$$f_k^{(5)}(\pi) = \left\{ -\frac{8}{\pi} k^4 + \frac{168}{\pi^3} k^2 \right\} \frac{\sin k\pi}{k\pi} \tag{39e}$$

$$f_k^{(6)}(\pi) = \left\{ -k^6 + \frac{72}{\pi^2} k^4 - \frac{1344}{\pi^4} k^2 \right\} \frac{\sin k\pi}{k\pi} \tag{39f}$$

$$f_k^{(7)}(\pi) = \left\{ \frac{12}{\pi} k^6 - \frac{696}{\pi^3} k^4 + \frac{12096}{\pi^5} k^2 \right\} \frac{\sin k\pi}{k\pi} \tag{39g}$$

where, after differentiating at the desired order, we have substituting everywhere

$$\cos k\pi = \left( \frac{1}{k\pi} - \frac{k\pi}{3} \right) \sin k\pi \tag{40}$$

which is nothing but a re-arrangement of equation (14) which defines $k$, which itself is nothing but the explicit form of the boundary condition (13). Then, the relations (38) follow by little algebra.

With the help of the large $k$ identity (A7) the relations (39) allows us to see that that for large $k$

$$f_k^{(2i+1)}(\pi) \sim f_k^{(2i)}(\pi) \sim (\pm) k^{2i-2} \tag{41}$$

a result that can be established in general by (33), (39a) which allows to deduce the large $k$ behavior of $f_k^{(2n)}(\pi)$ and (37) which shows that $f_k^{(2n+1)}(\pi) \sim f_k^{(2n)}(\pi)$ for large $k$ (modulo constant coefficients).

The importance of the relations (37) can hardly be overemphasized. Clearly, they show that the modes are only $C^2$ class periodic functions: $f_k(r)$ are even functions, whose odd derivatives (third and higher) do not vanish at $r=\pm\pi$ and therefore are not continuous. In infinite space, where the 'boundary' lies at infinity, relations such as (37) are void of content, as all quantities involved vanish. In compact space, these relations imply non trivial constraints on the fields $F$ and $H$ when combined with the regularity of the energy spectrum and energy transfer function.

Indeed, consider ideally regular energy spectra and energy transfer functions, that is, they are positive and exponentially decreasing for infinitely large $k$. Then, through the mode expansions (17) and (25) of the field $F$ and $H$, respectively, the relations (37) imply that

$$(\partial_r r^{-4} \partial_r r^4)^i \partial_r F(r)\Big|_{r=\pi} = 0, \qquad (\partial_r r^{-4} \partial_r r^4)^i H(r)\Big|_{r=\pi} = 0 \qquad (42)$$

for any value of $i$, as by (41) any derivative of the mode expansions converges for such spectral functions. Applying (37) we have used the fact that $\partial_r (r^{-4} \partial_r r^4 \partial_r)^i = (\partial_r r^{-4} \partial_r r^4)^i \partial_r$ which one may very easily verify by writing down explicitly the consecutive operators. These relations we shall call as the regularity conditions or relations, and the set of relations for $i \leq n$ form the order $n$, or the $n$-th order, set of regularity conditions.

For $n=\infty$, by (41) and the mode expansions of the fields $F$ and $H$, if the sum of $(\pm)k^{2i-2}E(k)$ and $(\pm)k^{2i-4}T(k)$ over all $k$, is finite for all $i$ then the regularity relations (42) hold for all $i$. If the spectral functions are ideally regular (positive and exponentially decreasing for large $k$) then the finiteness of these sums for any $i$ follows, and hence so do the relations (42). Thus (42) are a set of necessary relations for ideal regularity. There are then two intriguing facts. First, (42) are not adequate set of conditions for really well behaved energy spectrum and energy transfer functions, and second, in practice, only a finite number of these conditions can be employed in order to actually solve a problem in compact space. The point is though, that imposing the regularity relations in practice, we actually enforce regularity to an adequate degree so that these conditions are practically adequate: it turns out that by increasing the order of regularity the spectral functions remain physical for an increasingly larger regime of wavenumbers. Moreover, we shall see that the order of regularity is related to the 'well-resolved' scales, or equivalently and more concretely, to the number of well-defined energy moments which we need to have available in order to reconstruct in this language an actual case of DNS flows; those are always very few. These comments will become very clear in what follows.

We may give (42) an explicit form for the first few orders of regularity. Either through the relations (38) and the mode expansions of the fields, or by direct evaluation of (42), we obtain for $i=1,2,3$

$$F^{(3)}(\pi) = -\frac{4}{\pi} F''(\pi) \qquad (43a)$$

$$F^{(5)}(\pi) = -\frac{8}{\pi} F^{(4)}(\pi) + \frac{24}{\pi^3} F''(\pi) \qquad (43b)$$

$$F^{(7)}(\pi) = -\frac{12}{\pi} F^{(6)}(\pi) + \frac{168}{\pi^3} F^{(4)}(\pi) \qquad (43c)$$

for the field $F$ and

$$H''(\pi) = -\frac{4}{\pi} H'(\pi) \qquad (44a)$$

$$H^{(4)}(\pi) = -\frac{8}{\pi} H^{(3)}(\pi) + \frac{24}{\pi^3} H'(\pi) \tag{44b}$$

$$H^{(6)}(\pi) = -\frac{12}{\pi} H^{(5)}(\pi) + \frac{168}{\pi^3} H^{(3)}(\pi) \tag{44c}$$

for the field $H$. One should note that all regularity conditions, on both $F$ and $H$, express values of *odd* functions at $r=\pi$ in terms of even functions at that point. This is already clear at the level of the general relations (42).

We may now complete the discussion we started at the end of section IV A regarding the periodicity of the fields $F$ and $H$. Although the fields are $C^\infty$ differentiability class as functions over the *interval* $-\pi \leq r \leq \pi$, they can also be viewed as $C^2$ and $C^1$ periodic, respectively, over the *circle* parameterized by $-\pi \leq r \leq \pi$. Presumably, the field corresponding to the transverse autocorrelation function, which is related to $F$ by the first derivative relation $(r^2 F)'/2r$ [6], is also $C^1$ periodic. The 'failure' of having smoother periodicity is due to those derivatives of the fields which are odd functions and do not vanish at $r=\pm\pi$, hence becoming discontinuous as we cross the point $-\pi \sim +\pi$. That 'failure' or better, obstruction, is encoded fundamentally in the modes of the theory, expressed through the relations (37). The regularity conditions (42) are a formal statement of that obstruction at the level of the fields $F$ and $H$. The link between the modes and the fields is provided by the latter being admissible to physical interpretation.

**E. Interactions and renormalization**

Up to this point we have worked out the free field theory, that is, we constructed the modes of Karman-Howarth equation in compact space, and explored the host of the interesting facts about the nature of the fields $F$ and $H$ to which one is led to by the properties of those modes. But closing off the hierarchy at the level of one-dimensional fields, as discussed in section II, means that the field $H$ must be regarded as a functional of $F$ expressing its self-interaction. What makes things non-trivial is the regularity conditions (42). It is by no means obvious how and why any reasonable relationship between the fields should or could conform itself to those necessary conditions. It turns out that this difficulty resolves itself in a rather beautiful way.

**E1. A toy closure scheme**

Given a closure scheme $H=H[F]$ one derives an energy transfer $T[E]$. Generically that energy transfer ruins the regularity of the spectrum $E(k)$. A solution to this problem is to postulate closure schemes directly in spectral space, such that the energy transfer is well behaving. On the other hand, except for the fact such choices do seem somewhat unnatural, we are primarily interested in working with the Karman-Howarth equation and closure schemes directly in *r*-space. Making sense of closures in *r*-space unravels a fundamental aspect of our theory.

We need a simple, empirically admissible closure. Although detailed phenomenology is beyond the scope of this work, we need to use a closure scheme which maintains some of the basic phenomenological characteristics of isotropic turbulence. Also the model must be generalizable so that actual phenomenology can be done with it. An exceedingly elegant model of that kind is the model by Oberlack and Peters [7]. The model rests on the Kolmogorov's laws of isotropic turbulence [6]: in the inertial equilibrium range the (longitudinal) second and third order structure functions are given by specific scaling laws, the well-known two-thirds and four-fifths laws

$$\langle (u_l(\vec{r}) - u_l(0))^2 \rangle = C_2 (\varepsilon r)^{2/3}, \qquad \langle (u_l(\vec{r}) - u_l(0))^3 \rangle = -\frac{4}{5}\varepsilon r \qquad (45)$$

where $C_2 \sim 2$ is the famous Kolmogorov constant for the second order structure function. The model postulates a derivative relation between the third and second structure function such that the laws (45) are respected when the separation $r$ lies in the inertial subrange:

$$\langle (u_l(\vec{r}) - u_l(0))^3 \rangle = -\frac{4}{5 C_2^{3/2}} r \frac{\partial}{\partial r} \langle (u_l(\vec{r}) - u_l(0))^2 \rangle^{3/2} \qquad (46)$$

One verifies by inspection that indeed the model respects (45). With all its shortcomings this model is a good starting point of closures schemes in $r$-space. It is easy to see that the structure functions are related to the fields $F$ and $H$ by

$$\langle (u_l(\vec{r}) - u_l(0))^2 \rangle = 2F(0) - 2F(r), \qquad \langle (u_l(\vec{r}) - u_l(0))^3 \rangle = 6H(r) \qquad (47)$$

thus the model (46) becomes an explicit functional relationship $H = H[F]$:

$$H = -\alpha r \frac{\partial}{\partial r} [F(0) - F]^{3/2}, \qquad \alpha = \frac{4\sqrt{2}}{15 C_2^{3/2}} \qquad (48)$$

This closure is consistent with boundary conditions (9). Under the relation (48), $H(\pi) = 0$ follows from $\partial_r F(\pi) = 0$. That is, adopting (48) as a closure, the Karman-Howarth is rendered a closed dynamical equation for the field $F$ also in compact space, with the closure effecting the cascade process. Finally, one should note that the closure (48) respects the formal symmetry (28) of the Karman-Howarth equation. We shall return to this fact later on.

**E2. The difficulty made explicit**

Consider a regular spectrum of the form

$$E(k) = \tfrac{1}{24} k^4 e^{-k} \qquad (49)$$

which could have been used as an initialization spectrum in direct numerical simulation of turbulence. The wavenumbers $k$ here are 'our' $k$ defined by (14). The numerical coefficient is chosen such that the kinetic energy to be 1 (in suitable units) in the continuous case; here, the kinetic energy, which is given by (18), equals 0.9722. By formula (17) the spectrum (49) defines a field $F(r)$. Then, a field $H(r)$ is defined through the model (48), setting e.g. $C_2=2$, and from that the energy transfer $T(k)$ can be calculated by formula (24). The numerical calculation is straightforward using, for example, *Mathematica*, and the result is shown in Figure 1.

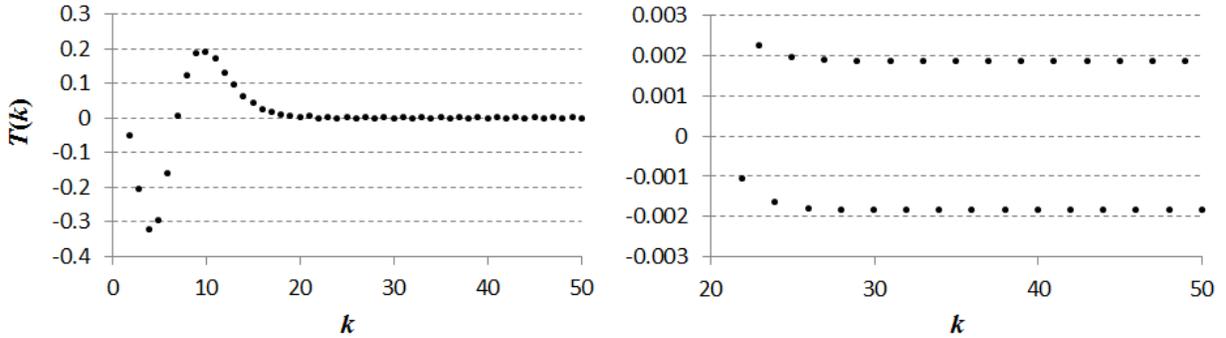

Figure 1: (a) The energy transfer function $T(k)$ as a function of $k$ derived from the closure (48) for the spectrum (49); (b) the asymptotic behavior of $T(k)$ for large wavenumbers, shown here by zooming to the regime of $k>20$.

The energy transfer $T(k)$ exhibits a physically acceptable form up to a certain wavenumber beyond which, instead of vanishing smoothly, it starts oscillating around zero while converging to an asymptotic behavior of the form

$$T(k) = \pm \text{constant} = (-1)^{[k]} \text{constant} \tag{50}$$

where $[k]$ is the integer part of $k$, as it shown in Figure 1b. This behavior persists for different regular spectra.

The problem of course is that, although the spectrum is regular, and therefore $F$ satisfies the regularity conditions (42) associated with it, the derived $H=H[F]$ does not satisfy the associated regularity condition in (42) and therefore does no satisfy any of relations (44). Indeed, checking (44a), we have that the closure (48) implies

$$H''(\pi) + \frac{4}{\pi} H'(\pi) = 3\alpha \sqrt{F(0) - F(\pi)} F''(\pi) \tag{51}$$

instead of being zero, as (44a) would require. The non-regular energy transfer will start transforming the spectrum to generic non-regular forms and the failure of satisfying any regularity condition will be permanent.

## E3. Renormalization

The regularity conditions not being satisfied under the closure (48) is the formal acknowledgement of the fact that the closure is not actually aware of the compactness of space. The closure is naturally constructed in unbounded space and still carries traces of this fact on it. The right hand side of (48) involves the field $F$, which will remain regular if it is allowed, and the external factor $r$. One may guess that this is the culprit, because in infinite space $r$ is a distance while in compact space is more like an angle, that ruins even the special periodicity properties of the fields $F$ and $H$ as they are encoded in the regularity conditions.

Indeed, let us write the closure in the form

$$H = -\alpha\, w(r) \frac{\partial}{\partial r}[F(0) - F]^{3/2} \tag{52}$$

where $w(r)$ is an odd, smooth and increasing function of $r$, such that $w(r)\sim r$ for distances quite smaller than the size $\pi$ of the compact space. This is always true for distances in the inertial subrange or smaller unless the Reynolds number is too small.

Regarding the boundary condition, the relation (52) implies first of all that $H(\pi) = 0$ still follows from $\partial_r F(\pi) = 0$. Now, following the terminology we introduced in the subsection D, given the first order regularity condition on $F$, condition (43a), we have that

$$H''(\pi) + \frac{4}{\pi} H'(\pi) = 3\alpha \sqrt{F(0) - F(\pi)}\, w'(\pi)\, F''(\pi) \tag{53}$$

That means that if

$$w'(\pi) = 0 \tag{54}$$

the first order regularity condition on $H$, condition (44a), is also satisfied. That is, regularity is completely satisfied at the first order.

The function $w(r)$ will be taken to be an (odd) polynomial of $r$, as a minimal choice, refraining from adding more structure to $w(r)$ that need not necessarily be there. The third order polynomial

$$w(r) = r - \frac{1}{3\pi^2} r^3 \tag{55}$$

clearly satisfies (54). Before proceeding to next orders of regularity, let us briefly pause to contemplate of what has been achieved here. First of all, all our results which were derived for periodicity with period $l=2\pi$, can be easily given for general $l$ if we re-scale everywhere the variable $r$ as $r \to (2\pi/l)\, r$. The function $w(r)$ then should be written as

$$w(r) = r - \frac{4}{3l^2} r^3 \tag{56}$$

bearing in mind that $w(r) \sim r$ for small $r$ and thus it should also be re-scaled same as $r$. Now if we take the limit $l \to \infty$, that is, *de-compactify* the theory, we obtain back the infinite space function $w(r)=r$. That is, as it is also clear from the first function in (55), the difference $w(r)–r$ involves terms that go away for large $l$. These terms, could be regarded as corrective terms, counter-terms, in the terminology of quantum field theory [10], which can be thought of as removing level by level the inconsistencies of the interactions with the basic principles of the theory. Although the analogy is quite vague, one cannot help but think that there is some correspondence. In quantum field theory, the inconsistencies amount to the arising of infinities in the calculation of physical quantities: They are removed by corrective terms which are not initially present but are allowed by the principles of theory, combined with the proper redefinition of fundamental physical quantities. This procedure is called renormalization. In the present theory, the inconsistencies are the irregularities of the spectrum and energy transfer. They are removed by corrective terms which are perfectly allowed by the construction of the closure scheme, specified by the application of regularity/periodicity conditions, through a proper redefinition the 'distance' factor $r$.

The most important common characteristic though, is that inconsistencies in both cases are removed *order by order* in perturbation theory, which also means at smaller and smaller distances, one step at a time. This is a fact in quantum field theory; regarding the present theory this last statement will become entirely clear just below and by the more detailed example presented in the next section. The order-parameter of perturbation theory in the present theory is the inverse square of compactification circumference $l$, $1/l^2$, which, by measuring $r$ through the turbulence microscales, essentially means that the order-parameter is the inverse Reynolds number. Finally, borrowing again terminology from quantum field theory, we may call (52) as the 'dressed' form of the closure/cascade term at the given order in the perturbation theory, as opposed to the 'bare' closure term given by (48).

We need to show that our construction does indeed work, and the most immediately available playground is the problem we set up in section E2. Before doing so, let us first derive $w(r)$ up to the third order of regularity, exploiting the equations (43) and (44).

Let us proceed to the application of the second order regularity conditions. We write $w(r)$ in the form of fifth order polynomial of $r$

$$w(r) = r - a_1 r^3 + a_2 r^5 \tag{57}$$

where the coefficients $a_1$, $a_2$ are to be determined. Given the second order regularity condition on $F$, relations (43a,b), the second order regularity condition on $H$, given by the relations (44a,b), are satisfied by

$$a_1 = \frac{2}{5\pi^2} \frac{1 - \frac{2}{15}\pi^2 \varphi''(\pi)}{1 - \frac{2}{75}\pi^2 \varphi''(\pi)}, \qquad a_2 = \frac{1}{25\pi^4} \frac{1 - \frac{2}{3}\pi^2 \varphi''(\pi)}{1 - \frac{2}{75}\pi^2 \varphi''(\pi)} \tag{58}$$

where we have defined the quantity $\varphi(r)$ by the formula

$$\varphi(r) = \frac{F(r)}{F(0) - F(\pi)} \tag{59}$$

which facilitates greatly the formal presentation of the coefficients of $w(r)$, at this order as well as at the higher orders. One observes that the coefficients $a$ are now functionals of the field $F$ carrying information about the field from the infinitesimal neighborhood of $\pi$. One observes that this dependence is such that the $w(r)$, and therefore the dressed closure, is still invariant under the formal symmetry (28). The importance of this fact will become clear later through our discussion in section IV.

At the third order we write

$$w(r) = r - a_1 r^3 + a_2 r^5 - a_3 r^7 \tag{60}$$

where again the coefficients are to be determined by the regularity conditions. Given the third order regularity conditions for the field $F$, equations (43), the third order conditions on $H$, equations (44), imply that

$$a_1 = \frac{11}{23\pi^2} \frac{1 - \frac{1042}{3465}\pi^2 \varphi''(\pi) + \frac{32}{693}\pi^4 (\varphi''(\pi))^2 + \frac{8}{693}\pi^4 \varphi^{(4)}(\pi)}{1 - \frac{38}{315}\pi^2 \varphi''(\pi) + \frac{40}{4347}\pi^4 (\varphi''(\pi))^2 + \frac{16}{7245}\pi^4 \varphi^{(4)}(\pi)} \tag{61a}$$

$$a_2 = \frac{13}{115\pi^4} \frac{1 - \frac{194}{273}\pi^2 \varphi''(\pi) + \frac{40}{273}\pi^4 (\varphi''(\pi))^2 + \frac{32}{819}\pi^4 \varphi^{(4)}(\pi)}{1 - \frac{38}{315}\pi^2 \varphi''(\pi) + \frac{40}{4347}\pi^4 (\varphi''(\pi))^2 + \frac{16}{7245}\pi^4 \varphi^{(4)}(\pi)} \tag{61b}$$

$$a_3 = \frac{3}{161\pi^6} \frac{1 - \frac{94}{135}\pi^2 \varphi''(\pi) + \frac{16}{81}\pi^4 (\varphi''(\pi))^2 + \frac{8}{135}\pi^4 \varphi^{(4)}(\pi)}{1 - \frac{38}{315}\pi^2 \varphi''(\pi) + \frac{40}{4347}\pi^4 (\varphi''(\pi))^2 + \frac{16}{7245}\pi^4 \varphi^{(4)}(\pi)} \tag{61c}$$

The dependence of the coefficients on $F$ is still given though the function $\varphi$, hence one may guess that that this a general rule and the symmetry (28) is respected in higher orders as well. The third order regularity function $w(r)$ is derived here mostly to illustrate the formal structure of the coefficients $a$ as we go to higher orders of regularity, as well as to use it in simple examples of theoretical interest. As we shall see in section IV, in applications of practical importance one hardly needs to impose regularity at that level.

We repeat the calculation for the energy transfer function $T(k)$ done in subsection E2 using now the dressed form of the closure for the first three orders of regularity. The results are shown in Figure 2. They are given in absolute value due to the sign-changing oscillations they exhibit once the first negative value is encountered. The result obtained in D2 is also included. The energy transfer remains regular for larger wavenumbers as we increase the order of regularity. One should also note that irregularity softens as we apply higher order regularity conditions, justifying explicitly their name. Indeed, the asymptotic behavior $T(k)=(-1)^{[k]}$ constant we found in section D2 by the bare cascade alone, takes the following forms (at the indicated order of regularity)

$$T_{1-\text{order}}(k) = \frac{(-1)^{[k]}}{k^2}, \qquad T_{2-\text{order}}(k) = \frac{(-1)^{[k]}}{k^4}, \qquad T_{3-\text{order}}(k) = \frac{(-1)^{[k]}}{k^6} \qquad (62)$$

up to a proportionality constant. The results are of course numerical but the faith in them will be strengthened by our arguments given below as well as in the next sections.

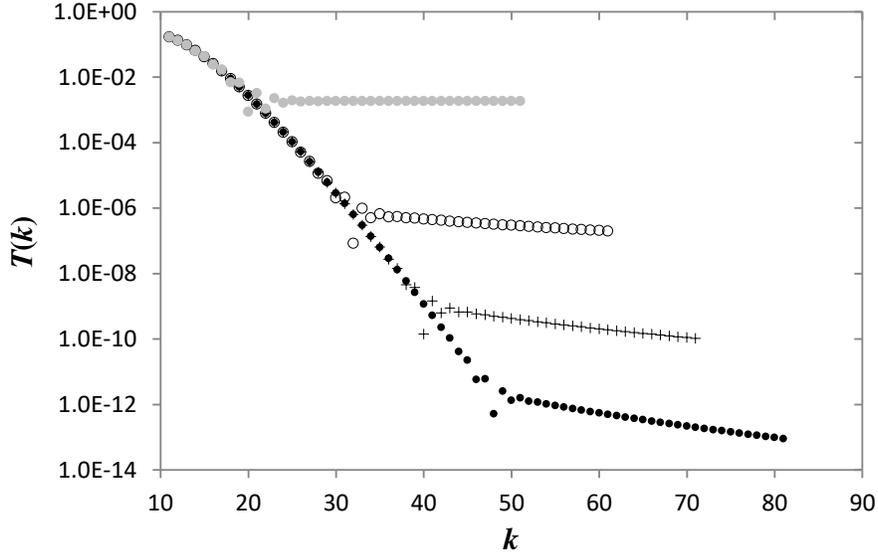

Figure 2: Plots of the absolute values of the energy transfer $T(k)$ for the zeroth, first, second and third order of regularity (the order increases downwards). $T(k)$ becomes regular for increasingly higher wavenumbers as we impose higher order regularity conditions on the closure. The softening of the irregularities advocated by (62) is also visible.

We observe that by increasing the order of the polynomial $w(r)$ by two, so that to satisfy the next order regularity, the regularity of $T(k)$ increases also by two orders, as quantified by its asymptotic scaling. That is, each new counter-term, which contains an additional factor of $r^2$, removes a factor of $k^{-2}$ from the leading irregularity. Dimension-wise this makes much sense. Nonetheless, we may derive analytically these asymptotics given in (62) by looking at simple examples of $H(r)$ such that the regularity conditions of increasing order are satisfied for that field.

The simplest examples we may use are polynomials. Such functions allow explicit calculation of the energy transfer in compact space and have enough content to exhibit analytically the asymptotic behavior given in (62). The field $H(r)$ is a negative, odd function, with no linear term in its expansion around $r=0$. Hence consider first for example

$$H(r) = -r^3 + \frac{r^5}{\pi} \tag{63}$$

where an overall positive constant coefficient is left understood. This field satisfies just the boundary condition, equation (9): $H(\pi)=0$. Then by equation (24) we may calculate the energy transfer explicitly. Using the relation (40), which implements the defining relation of the $k$'s, and the asymptotic relation (A7) one finds quite simply and explicitly that for large $k$

$$T(k) = 22\pi^2 (-1)^{[k]} \tag{64}$$

which is the asymptotic behavior (50), modulo a proportionality constant, found by the bare form (48) of the cascade.

Next consider the example

$$H(r) = -r^3 + \frac{26 r^5}{15\pi^2} - \frac{11 r^7}{15\pi^4} \tag{65}$$

which is constructed to satisfy the first order regularity condition (44a), and of course the boundary condition $H(\pi)=0$. Then, by the same steps as above, we calculate explicitly the large $k$ behavior of the energy transfer associated with this field to find

$$T(k) = \frac{4344}{5} \frac{(-1)^{[k]}}{k^2} \tag{66}$$

that is, exactly the asymptotic behavior we found numerically at the level of the first order regularity, quoted in (62). In a similar manner one verifies by examples the asymptotics for the higher orders given in (62).

Similarly we may derive the asymptotics for the energy spectrum. The field $F(r)$ is an even function which satisfies the boundary condition $F'(\pi)=0$. Given simple forms, such as even order polynomials, the associated spectrum is calculated explicitly by equation (21), with the aid of (40) and (A7) as above. Then it is easy to show that

$$E_{\text{bare}}(k) = \frac{(-1)^{[k]}}{k^2}, \quad E_{1-\text{order}}(k) = \frac{(-1)^{[k]}}{k^4}, \quad E_{2-\text{order}}(k) = \frac{(-1)^{[k]}}{k^6}, \quad E_{3-\text{order}}(k) = \frac{(-1)^{[k]}}{k^8} \tag{67}$$

always up to a proportionality constant, when the field *F* satisfies regularity to the indicated order, respectively. The first result is obtained when no regularity condition is imposed on *F*. Note that, recalling the discussion of subsection D, the regularity conditions of order *n* implies that the sums $(\pm)k^{2n-2}E(k)$ and $(\pm)k^{2n-4}T(k)$ over all *k*, converge, although the large *k* behavior of the spectral functions it was not known. The results obtained here show that these sums indeed converge as the summand in both case behaves as a safe $k^{-4}$.

By this elementary analysis we obtained results which can also be derived, with a bit more effort, through more complicated functions, e.g. rational functions, which contain a full power series of *r*. With such functions, as we increase the order of regularity and the irregularities of the spectrum are softened and pushed deeper in the dissipation subrange, the usual exponential fall-off of the spectrum is exhibited for the large *k* for which the spectrum remains physical. We will see this behavior in section VI where these ideas are tested, and the renormalized closure operates within the Karman-Howarth equation so that the increased regularity of the field F and therefore of the spectrum are produced as a *result*.

Finally, contrasting (67) with the results for the energy transfer in (50) and (62) we see that the asymptotic i.e. the dominant irregularities of the *E*(*k*) and *T*(*k*) are related by

$$E(k) \text{ proportional to } \frac{T(k)}{k^2} \tag{68}$$

a relation which will re-appear in the next section and it will be understood as an immediate result of the energy balance equation (23) of turbulence in steady state.

**E4. Regularity, moments and scale**

The number of energy moments which are defined in the sense of (32), depends on the order of regularity. In turn, the order of regularity determines the scales where the spectrum remains physical, which in this context we have defined as meaning positive and decreasing (bearing in mind that irregularities arise characteristically in the dissipation subrange). Nonetheless, the analyticity of the field *F* dictates and allows all energy moments to be defined in a weak sense.

To clarify these statements we analyze in some more detail what the asymptotics (67) imply about the energy moments and the field *F*. The phenomena of interest are revealed with greater ease by working with continuous wavenumbers, hence we will do so. It is very instructive to consider first energy spectra with power law asymptotics which are *not* oscillatory. Consider then a spectrum *E*(*k*) which behaves as $E(k) \sim 1/k^2$ for large *k*. Clearly, for such a spectrum, no other energy moment is defined but the total energy. As we discussed in section IV D the energy moments are related to the derivatives of *F* at *r*=0, therefore we should also look at the small *r* behavior of *F*. The small *r* behavior of the field *F* can be deduced by using the formula (10) restricting the lower end-point of *k* to some wavenumber $k_0$. The integration can be done explicitly, though only the small *r* behavior of the result is important. One finds

$$\frac{2}{3k_0} - \frac{\pi}{8}|r| + \frac{k_0}{15}r^2 + O(r^4) \tag{69}$$

One may immediately see the problem. Due to the term involving the absolute value of $r$ the field $F$ is not differentiable at $r=0$. This is expected as, in order to define at least the dissipation $\varepsilon$, we need the field $F$ to be twice differentiable at $r=0$. An analogous situation continues for softer asymptotics (which can be regarded as sub-dominant terms of the same spectrum). Consider a spectrum $E(k)$ which behaves as $E(k) \sim 1/k^4$ for large $k$. Then the small $r$ behavior of $F$ coming from that reads

$$\frac{2}{9k_0^3} - \frac{1}{15k_0}r^2 + \frac{\pi}{144}|r|^3 + \frac{k_0}{420}r^4 + O(r^5) \tag{70}$$

This $F$ is only twice differentiable, as expected according to the discussion of section IV D, by the fact that only the dissipation ($i=1$ moment) and not the palinstrophy ($i=2$ moment) can be defined for that (part of the) spectrum.

In the case of asymptotically *oscillatory* spectrum things are different. Indeed, such a spectrum is unphysical beyond a certain wavenumber. On the other hand, the energy moments for such a spectrum are well-defined in a weak (regularized) sense, and this is enough to guaranty differentiability of the field $F$ at $r=0$. Consider an asymptotically oscillatory spectrum $E(k)$ which behaves, say, as $E(k) \sim \sin(k)/k^2$ for large $k$. The small $r$ behavior of the field $F$ associated with this asymptotics is deduced again through (10) with a lower end-point $k_0$. This time the result, which is explicit though too cumbersome to quote here, is a perfectly analytic function of $r$ i.e., all its derivatives at $r=0$ exists. On the other hand, the spectrum asymptotics imply apparently that the total energy is the only energy moment which is well-defined. This contradicts our discussion in section IV D.

The resolution of this little puzzle lies precisely in the oscillatory nature of the spectrum. The energy moment integrals (for the given spectrum) depend on the convergence of the integrals

$$\int^\infty k^{2i-2} \sin k \, dk \tag{71}$$

$i=1,2,..$ , which apparently do not converge; nonetheless, defining the moments in the weak sense

$$\lim_{a \to 0} \int_0^\infty e^{-ak} k^{2i} E(k) dk \tag{72}$$

their existence depends on the integrals

$$\lim_{a\to 0}\int_0^\infty e^{-ak}k^{2i-2}\sin k\,dk \tag{73}$$

which are finite for all *i*. The reason for this phenomenon is the oscillations: introducing the regularization factor $e^{-ak}$ we control the size of the large contributions to the integral, coming from the large *k*, which turn out to cancel each other as we let the factor go away. Such procedures are common place in quantum field theory [10] and are also in use in theory of turbulence [6]. Thus the moments do exist in a weak sense.

From the point of view of the field *F*, one should first of all bear in mind that the derivatives of *F* 'at' $r=0$ are also defined in a weak sense: by the very form of the Karman-Howarth equation (4), with those factors of *r* in the denominator, the information about the point $r=0$ must always be understood though the limit $r\to 0$. In particular, a spectrum whose asymptotics of an oscillatory power-law form, such as $E(k)\sim\sin(k)/k^2$, produce analytic $F(r)$ in the sense of functions such as $\sin(r)/r$, where the point $r=0$ is added by continuity. Nonetheless, these functions *are* analytic which allows for the following manipulation. The relation between the derivatives of *F* at $r=0$ and the energy moments, given by the limit $r\to 0$ of equation (34), is derived by sum or integrating over all *k* the equation (33). If we multiply with the regularizing factor $e^{-ak}$ before summing or integrating one obtains

$$\tfrac{3}{2}(-r^{-4}\partial_r r^4\partial_r)^i\{\tfrac{2}{3}\sum_k e^{-ak}E(k)f_k(r)\} = \sum_k e^{-ak}k^{2i}E(k)f_k(r) \tag{74}$$

Now in the limit $a\to 0$ the bracketed quantity in the left hand side does give us $F(r)$ which is a well-defined analytic function, that is, the limit can be explicitly performed on the left hand side. Taking $r\to 0$, the limit can be taken explicitly on the left hand side, which is simplified by $f_k(r)\to 1$, to obtain

$$\lim_{r\to 0}\{\tfrac{3}{2}(-r^{-4}\partial_r r^4\partial_r)^i F(r)\} = \lim_{a\to 0}\sum_k e^{-ak}k^{2i}E(k) \tag{75}$$

It is this form of the relation between the energy moments and the derivatives of *F* which is realized here. In fact, one may think backwards and realize the following: an analytic *F* is in general associated with energy spectra whose moments are at least defined in the weak sense.

Of course such spectra must become unphysical beyond a certain wavenumber. The point is though that where this happens can be controlled and pushed – also controllably – to higher wavenumbers by the regularity conditions, as discussed in the previous sections. Let us be explicit. Recall the energy spectrum asymptotics (67) for the various orders of regularity. We start by observing that the bare result allows us to define the total energy in the strong (i.e. the usual) sense, while the dissipation and the higher energy moments are ill-defined in the strong sense and they are well-defined in the weak sense. The dissipation spectrum $k^2 E(k)$ approaches $(-1)^{[k]}$, which is a mild non-convergence, and can be regarded as marginally ill-defined. At the first order of regularity the energy and dissipation are well-defined in the strong sense while the higher moments are defined in the weak sense. The palinstrophy is marginally ill-defined as its spectrum $k^2E(k)$ approaches $(-1)^{[k]}$. At the second order of the moments up and including the palinstrophy are well-defined in the strong sense and the $i \geq 3$ moments are defined in the weak sense. The $i=3$ moment, which we shall call below as the $k^6$-moment, is marginally ill-defined. One can rather safely guess that this pattern continues to higher orders. In practice, where we work with a finite number of wavenumbers, say $k_1,\ldots,k_N$, the regularization of the sums in (75) is very simply implemented by averaging suitably partial sums as explained in Appendix F. This is adequate to ensure (75) in many significant figures, as we shall see in section VI.

During the process of applying higher orders of regularity we push the point where the spectra become unphysical deeper into the dissipation subrange. The physical regime is quantified more naturally in terms of the dimensionless wavenumber $k\eta$ (where $\eta=(v^3/\varepsilon)^{1/4}$ is the Kolmogorov length scale of the dissipation subrange). Irregularities of the spectrum start beyond the peak the of the dissipation spectrum, i.e. within the dissipation subrange, although, strictly speaking, weak irregularities regarding the monotonicity alone, may start already at the smallest wavenumbers as we increase the Reynolds number. We shall see in section VI that the physical regime of the spectrum is extended in units of $k\eta$ every time we apply regularity at a higher order. The number of energy moments defined in the strong sense increases order by order because the stronger i.e., larger, irregularities are removed allowing the underlying physical spectrum to reveal itself to higher wavenumbers. Thus the order of regularity of the field $F$ is associated with the scale (quantified by $k\eta$) the energy spectrum remains physical. That scale is a strong analogue of the resolution scale of the direct numerical simulations: at the $i$-th order of regularity, at which the $i$-th energy moment is well-defined in the strong sense, the spectrum remains physical far deeper in the dissipation range than the resolution required usually in order to obtain a practically acceptable estimate of that moment. Given that the resolutions of the current direct numerical simulations is given by very few $k\eta$ [4][2] and hence very few moments can be acceptably estimated we understand that only very low order regularity needs to be imposed in order to obtain practically useful physical spectra.

## V. COMPACT SPACE, THE DNS AND LARGE SCALE CUTOFF

We should recall at this point that our mathematical efforts were motivated by the characteristics of the DNS flows, and they are an attempt to formally reconcile compactness of physical space with exact isotropy, as we discussed in section III. We have seen that this can certainly be done, somehow, by moving at the level of the correlation function scalars and work there. On the other hand it is not clear whether our construction has anything to do with the DNS flows, or for that matter, with flows at all. A priori it is not clear whether there exist underlying velocity fields whose fluctuations are statistically described by our correlation functions. Nonetheless the properties of these quantities reflect important features of analogous quantities in the DNS flows. In fact, there is a close relationship between the quantities of the compactified theory and analogous quantities in ideal (i.e. infinite space) isotropic flows with an explicit cutoff in their low wavenumbers. Such a cutoff is interpreted as an explicit upper bound in the length scales and the 'size' of the modes of the flow. Though the claimed relationship is intuitively rather clear it is illuminating to make the details explicit.

### A. The spectral space of the DNS flows

The spectral space of the DNS flows consists in vectors which are triplets of integers, $k=(n_1,n_2,n_3)$. The spectral space is organized in shells of wavenumbers labeled by integers $n$ such that $n–0.5<|k|<n+0.5$. Averaging functions of all wavenumbers in each shell one obtains direction-independent quantities which are regarded as functions of $n$. The larger the (shell) number $n$, the better the approximation becomes. Then one may think of each shell labeled by $n$ as a single wavenumber $n$. We have discussed that the $k$'s of the compact space, given by the solutions of (14), approach integers values as they become large. Thus for large wavenumbers the two spectral spaces become similar, $k \simeq n$. The dimensionful compact wavenumbers, $2\pi k/l$, also coincide if we choose a common length $l$ entering in the boundary conditions (8) and (9). This is the reason why we have chosen the circumference $l$ of the compact space to coincide with the periodicity $2\pi$ of the DNS.

For the relatively small $n$ things become more intricate. Especially as we approach the smaller values that correspond to the increasingly non-isotropic modes, the shell approximation of the DNS flows spectral space becomes a very crude one: There are too few wavenumber vectors in each shell, and the use of shells serves more as completing the bookkeeping of each spectral function in question in this one-dimensional reorganization of spectral space. On the compactified theory side, the spectral space is inherently one-dimensional and every single mode is as isotropic as any other, as they all obey the equations of the ideal isotropic flows. What *is* different in the relatively small values of $k$ is that they lie slightly away from the nearest integer, as we see in (15). There is a decoupled $k = 0$ mode that contributes pure constants in the field $F$ and the lowest non-zero wavenumber $k$ is near the integer 2 and not near 1. One may interpret near-integer wavenumbers as a more carefully weighted labeling of the shells in the spectral space of the DNS flows, which become increasingly sparse for the longer waves. On the downside, at the lowest end of the two spectra the correspondence between them becomes nebulous. A good reason for that is that the lowest modes, and especially the $|k|=1$ ones, are the most anisotropic modes of the DNS flows.

### B. Large scale cutoff

The vanishing (compact space) Loitsyansky integral arose in our discussion above as a fundamental property of the compactified theory. In section IV C we learned that a major implication of this fact is that the conserved Loitsyansky integral of the theory does not operate the same way as in ideal isotropic flows: scale-wise our system has only to conform to the compactness of the domain. In this respect it resembles the DNS flows which have no other obvious constraint scale-wise but the compactness of the space they live. Additionally, the Loitsyansky integral of the compact space is essentially the energy of the zero-mode of the flow, which is not dynamical, and serves only as a constant bottom of the energy $K$ which may as well set to zero. The seemingly peculiar implication of zero-Loitsyansky condition (31) is that $F$, which is understood as an analogue to the *longitudinal* autocorrelation functions, must become negative somewhere in its domain. We argue that this is a natural byproduct of the behavior of actual flows that are bounded, explicitly or effectively, in space.

We start with the ideal isotropic flows which live in infinite space. Formula (10) relates the spectrum $E(k)$ with the longitudinal autocorrelation function $F(r)$. Consider a simple spectrum with an infrared cutoff: $E(k)=0$ for $k<k_0$ for some wavenumber $k_0$ while $E(k)$ assumes any reasonable form for the higher wavenumbers. To be concrete, let us take for example that $E(k)=k^4 e^{-k}$ for $k>k_0$, in suitable units. Introducing a sharp cutoff is a legitimate way to allow modes only up to a certain size within the infinite space; information of this kind is stored more consistently in spectral space where the positivity of the spectrum is a clear cut condition that needs to be satisfied with no analog in physical space. It is then straightforward to show that the asymptotic behavior of $F(r)$ at large distances is

$$F(r) = -2k_0^2 e^{-k_0} \frac{\sin k_0 r}{r^3} + 2k_0(k_0 - 3)e^{-k_0} \frac{\cos k_0 r}{r^4} + O(r^{-5} \sin k_0 r) \tag{76}$$

Clearly, this $F(r)$ vanishes, becomes negative and starts oscillating around zero for distances beyond $r=O(1/k_0)$. The same happens when the sharp infrared cutoff is implemented in a smooth way by some exponential function: the result exhibits the same behavior, only the calculations are a bit harder.

In the brief discussion that follows we shall describe $F(r)$ such as (76) as asymptotically oscillating. Their trademark properties are the existence of extrema that are both positive and negative i.e. the existence of nodes between them. The usual longitudinal functions of isotropic turbulence do not exhibit such behavior. There is a close relationship between the field $F$ of the compactified theory and these asymptotically oscillating functions. By the boundary conditions (9b) each of the end points of the interval $-\pi \leq r \leq \pi$ is a stationary point of the field $F$ (and as usually we use the even-ness of $F$ to restrict ourselves to the half of that interval). In the simplest configurations the field $F$ is continuously decreasing and reaches its first minimum at the end of the interval. Now the field $F$, modulo an ignorable additive constant, satisfies (31). That means that in some neighborhood of its minimum the field $F$ must have passed to the negative values. That is, there must be a node between $r = 0$ and $r = \pi$. These configurations of the field $F$ can be regarded as appropriate asymptotically oscillating configurations truncated at their first minimum. This is the form of the first mode, that is, the eigenfunction (11) which corresponds to the lowest non-zero wavenumber $k=1.835$ [given in (15)].

Similarly, configurations that possess multiple nodes, that is, two or more nodes in the interval $0 \leq r \leq \pi$, can be regarded as asymptotically oscillating ones truncated at an appropriate minimum of maximum. Now, the form of the configurations is only half of the story. The discrete spectral space of the compactified theory introduces an a priori cutoff on the energy spectrum of any configuration of the field $F$. That cutoff can be regarded as a counterpart of the sharp spectral cutoff that gives rise to the asymptotically oscillating configurations. But there is an important difference between them: the cutoff of the compactified theory is permanent as compactness is permanent. The single- and multi-node configurations can be distinguished on the basis of this difference.

Single-node configurations of the field $F$ arise naturally. Plausible, positive-definite spectra that one uses to set up initial conditions for the field $F$ are most likely to lead to configurations with a single node. As noted already above, the node may move but not disappear, courtesy of (31). From the spectral space point of view, the permanence of these configurations rests on the naturalness of their energy spectra. The spectra possess a single cutoff which is the a priori and *permanent* spectral cutoff of the compactified theory. Multi-node configurations arise when we consider spectra with an additional low wavenumber cutoff. That is, by repeating on the discrete spectral space the procedure we applied in the continuous spectral space of the ideal isotropic flows that led to the asymptotically oscillating functions. In a sense, the multi-node configurations are a counterpart of the asymptotically oscillating configurations in compact space. In their respective domains both kinds of configurations have at least one node too many, or equivalently, they have one sharp cutoff too many in the size of their modes rendering them unstable structures. The reason is that the sharp cutoff in their spectra is unsustainable. The nonlinear interactions are continuously redistributing energy into wavenumbers in the vicinity of the cutoff softening its sharpness. Gradually the nodes are untangled and the configurations settle into the most elementary forms they assume in their respective domains. In compact space, that form is the single-node configuration. In conclusion, thinking in analogy with the ideal isotropic flows we acquired an intuitive basis for the necessary appearance of nodes in the configurations of $F$, a fact that is elegantly encoded in (31), and clarified their function.

## VI. FORCED TURBULENCE IN COMPACT SPACE

### A. Forcing in compact space

Assume that we feed each mode $k$ with energy at a rate $\varepsilon_k$. That means that the (total) energy balance (2) reads

$$\frac{dK}{dt} = -\varepsilon + \sum_k \varepsilon_k \tag{77}$$

while the spectral form of the energy balance (23) reads

$$\frac{dE(k)}{dt} = -2\nu k^2 E(k) + T(k) + \varepsilon_k \tag{78}$$

which, in *r*-space, takes the form of a Karman-Howarth equation (4) with forcing terms

$$\frac{\partial F}{\partial t} = \frac{1}{r^4}\frac{\partial}{\partial r}\left[r^4\left(2\nu\frac{\partial F}{\partial r} + H\right)\right] + \frac{2}{3}\sum_k \varepsilon_k f_k(r) \tag{79}$$

The numerical coefficient 2/3, present also in the mode expansion (17), reflects the normalization of the field *F* with respect to the total energy. Each mode *k* is represented in *r*-space simply and directly by the corresponding eigenfunction $f_k(r)$. This allows the simple representation of forcing we see in (79). The formulas (17), (19) and (21) allow the back and forth transformation between (78) and (79).

In the DNS, forcing is usually applied in the first lowest shells, that is, bunches of DNS turbulence modes centered around *k*~1 and *k*~2, as we explained in section V A, such that the total energy *K* is kept constant. Here these two shells are represented by the first wavenumber given in (15), $k_1$=1.835. Therefore we feed energy only the first mode, so that the total energy is kept constant. Then (77) tells us that the forcing rate on the first mode $k_1$ must be equal to the dissipation: $\varepsilon_{k_1} = \varepsilon$, where $\varepsilon$ is given at any instant of time by equation (6). In all we have

$$\frac{\partial F}{\partial t} = \frac{1}{r^4}\frac{\partial}{\partial r}\left[r^4\left(2\nu\frac{\partial F}{\partial r} + H\right)\right] + \frac{2}{3}\varepsilon f_{k_1}(r) \tag{80}$$

Given some initial configuration of the field *F* which corresponds to a regular spectrum, a closure scheme *H*[*F*], as we explained in sections IV D and IV E, and the boundary conditions on *F* given by (9) we have a well-defined problem to solve.

We should also make sure that the initial configuration of *F* respects the condition (31), of zero Loitsyansky integral in compact space. Indeed, as discussed in section IV B, the Loitsyansky integral in compact space is nothing but the energy *E*(0) of the zero-mode of spectrum, given explicitly by equation (22). As discussed in IV B, the zero-mode is entirely decoupled from all the *k*>0 modes and its energy *E*(0) is a constant of motion, acting merely as an arbitrary bottom of the total energy without physical relevance. The existence and the non-dynamical nature of the zero-mode is an implication of the formal symmetry (28) of the Karman-Howarth equation with the boundary conditions (9); this symmetry is respected by the dressed closure scheme (52), as discussed in section IV E3, and hence by (79). Thus, if we want the sum of *E*(*k*)'s to correspond to the actual total energy *K* of the dynamical degrees of freedom, that is, if we want to be able to identify *F*(0) with (2/3)*K* according to equation (7), then we must choose initial configurations such that *E*(0)=0, which will be preserved throughout the evolution of the system by the constancy of $E(0)$. The condition $E(0)=0$ is equivalent to the integral condition (31), which we re-write in the half interval, utilizing the evenness of the field $F$ as usually, requiring that

$$\int_0^\pi r^4 F(r) dr = 0 \tag{81}$$

Once the initial configuration of $F$ is consistent with (81) we have a complete formulation of the problem.

**B Steady state**

Same as in the DNS, we are primarily interested in the asymptotic state of the system, which is a *steady state*: there is no time dependence on any quantity. Then

$$0 = \frac{1}{r^4}\frac{\partial}{\partial r}\left[r^4\left(2\nu\frac{\partial F}{\partial r} + H\right)\right] + \frac{2}{3}\varepsilon f_{k_1}(r) \tag{82}$$

We may multiply this equation by $r^4$ and integrate from $r=0$ to $r=r$. As both $\partial_r F$ and $H$ vanish at $r=0$ we obtain the *first order* equation

$$r^4(2\nu F' + H) + \frac{2}{3}\varepsilon \int_0^r r^4 f_{k_1}(r) dr = 0 \tag{83}$$

where we denote differentiation with respect to $r$ by a prime. The integral which arose in the forcing term can be understood as an incomplete form of the compact space Loitsyansky integral of the mode $k_1$. By equation (C2) it is related to the derivative of the corresponding mode $k_1$. Thus, we obtain the rather elegant *steady state equation*

$$2\nu F' + H - \frac{2}{3}\frac{\varepsilon}{k_1^2} f'_{k_1}(r) = 0 \tag{84}$$

At $r=\pi$, equation (84) is satisfied due to the boundary conditions (9) on the fields $F$ and $H$ and (13) for the eigenfunction, i.e., each term vanishes independently. Clearly, the same procedure can be also applied to the general equation (79). Thus, a DNS type of forcing can be nicely accommodated in our all-isotropic formalism leading to a first order differential equations for the steady state (as long as the closure is also first order in $F$).

An important implication is that if the field $F$ is regular up to a certain order then (84) implies that we must have a closure $H[F]$ such that $H$ is also regular at that order. Indeed, assume e.g. that $F$ is third order regular, that is, conditions (43) hold. As these follow from the relations (38) of the eigenfunctions, similar relations hold for the forcing term in (84). Therefore, (81) *requires us* to have a field *H* which is also third order regular. In section IV E3 we discussed how this accomplished by renormalizing the selected closure at the desired order.

Let us write down (84) in an explicit form, using also the renormalized form of the closure given in equation (52),

$$2\nu F'(r) + \frac{3}{2}\alpha w(r)\sqrt{u'^2 - F(r)}F'(r) - \frac{2}{3}\frac{\varepsilon}{k_1^2}f'_{k_1}(r) = 0 \qquad (85)$$

We may immediately observe a few things. First of all, the forcing term is proportional to an eigenfunction and hence vanishes at $r=\pi$ by the boundary condition which is obeyed by the eigenfunctions and the field *F*, relation (13) and (9) respectively. Then, by the form of the cascade process term, the derivative of the field *F* is a common factor in the first two terms of (84). Thus, the boundary condition $F'(\pi)=0$ is actually implied by (82) and need not be imposed independently. This makes sense, as all we are left with in the steady state is the first order equation (84), whose only necessary 'initial data' along the line of *r*, are exhausted by the value of the field *F* at $r=0$, that is, by the value of $u'^2=(2/3)K$, which is set once and for all by the initial condition and preserved by the forcing.

Secondly, in order to actually being in able to identify $F(0)= u'^2$, equation (85) must be supplemented by the zero Loitsyansky integral condition (81). This condition is imposed at the initial configurations on *F* and is carried over time by the conservation of the compact space Loitsyansky integral all the way to the steady state. Working formally directly with the steady state, we must impose it independently in order to obtain the unique solution of (85) that corresponds to initial configurations of zero Loitsyansky integral.

In order to immediately see the role of this condition, imagine that somehow the cascade process is switched off. Then equation (85) is solved trivially to give

$$F(r) = \frac{\varepsilon}{3\nu k_1^2}f_{k_1}(r) \qquad (86)$$

*modulo* an additive integration constant. But the zero Loitsyansky integral condition (81) and the fact that the mode satisfies already that condition, equation (20), we have that the additive constant must be *zero*. Then, using the fact that $F(0)= u'^2$ and $f_{k1}(0)=1$ we have that

$$\varepsilon = 3\nu k_1^2 u'^2 \qquad (87)$$

that is we obtain the *resulting* value of dissipation, given the viscosity, the forcing mode and the initial energy. Although the steady state value of the dissipation of the full problem (85) is not given by (87), the latter does emphasize the correspondence between the initially given energy and the steady state value of the dissipation. Clearly, without employing the condition (81) we would not be able to identify the actual total energy of the system and derive a unique energy balance relation such as (87).

It is convenient to introduce the normalized (and dimensionless) field

$$f(r) = \frac{F(r)}{F(0)} \tag{88}$$

which satisfies $f(0) = 1$. We re-write (85) into the form

$$f'(r) + \frac{3\alpha}{4}\frac{u'}{v} w(r)\sqrt{1-f(r)} f'(r) - \frac{\varepsilon}{3vk_1^2 u'^2} f'_{k_1}(r) = 0 \tag{89}$$

This equation involves two dimensionless constants. The quantity $u'/v$ is a dimensionless quantity, in the arbitrary length units we use to measure $r$. It is essentially a Reynolds number of the general form $LU/v$, where $L$ and $U$ are associated with size and the total energy of the system respectively. Given that the forcing maintains the energy $(3/2)u'^2$ constant at its initial value, this quantity is fixed by the initial conditions. Also, recalling the Taylor microscale Reynolds number [6]

$$\text{Re}_\lambda = \sqrt{\frac{15 u'^2}{\varepsilon v}} \tag{90}$$

we obtain a dimensionless quantity associated with the dissipation of the steady state. Therefore this quantity characterizes of the steady state we are lead to, for the given initial conditions. It easy to see the coefficient of the forcing term reads $5(u'/v)k_1^{-2}\text{Re}_\lambda^{-2}$, that is, the coefficients of (89) are indeed expressed in terms of these two Reynolds numbers. Solving the problem means that we determine $\varepsilon$ for the given $u'$. This means, in turn, that we determine $\text{Re}_\lambda$ for the given $u'/v$. We should perhaps note that for a given initial configuration, say $F_0(r)$, there is also an initial value of the dissipation, say $\varepsilon_0$, given by $-15vF_0''(0)$. But dissipation is not preserved by the forcing, therefore $\varepsilon_0$ is not a characteristic of the initial state that remains relevant in the steady state.

## C Renormalization

The function $w(r)$ is in general a functional of the field $f$, which we constructed in section IV E3 for the first three orders of regularity. Indeed, beyond the first order the function $w(r)$ depends on the derivatives of the field $F$ and therefore of $f$ at $r=\pi$. This is slightly unsettling, as higher order derivatives of the unknown function enter the (first order) differential equation (89). On the other hand, $w(r)$ is a *local* functional of $f$ at the point $r=\pi$. Therefore we may use the information contained in the equation (89) in the neighborhood of $r=\pi$ to re-write $w(r)$ in an equivalent but more agreeable way such that to depend only on the *value* of the field $f$ at $r=\pi$. The function $\varphi(r)$, defined in (59), reads equivalently

$$\varphi(r) = \frac{f(r)}{1 - f(\pi)} \tag{91}$$

Consider the $w(r)$ imposed by the second order regularity conditions. Its coefficients, given by equation (58), involve the second derivative of $f$ at $r=\pi$. Differentiating (89) once and evaluating the result at $r=\pi$, using also the boundary condition $f'(\pi) = 0$, we obtain a quadratic equation for $f''(\pi) = 0$ which is easily solved to obtain

$$f''(\pi) = \frac{75(1 - f_\pi)}{4\pi^2} \frac{1 + A + B - \sqrt{(1 + A + B)^2 - 4AB}}{A} \tag{92}$$

where

$$A = \frac{25\nu}{12\pi u' \alpha \sqrt{1 - f_\pi}}, \quad B = -\frac{\varepsilon \sin k_1 \pi}{54 k_1 u'^3 \alpha (1 - f_\pi)^{3/2}} \tag{93}$$

Note that we write $f_\pi$ in place of $f(\pi)$. We also used (39b) to replace the quantity $f''_{k_1}(\pi)$ which arises. Note that both constants $A$ and $B$ are positive. Using the result (92) into (58), we find

$$a_1 = \frac{4(1 - 2\delta)}{5\pi^2}, \quad a_2 = \frac{(1 - 12\delta)}{25\pi^4}, \quad \delta = \sqrt{(1 + A + B)^2 - 4AB} - 1 + B - A \tag{94}$$

Thus, $w(r)$ is written in such a way to involve only the *value* of $f$ at $r=\pi$. This is quite convenient for the numerical solution of the problem.

As a matter a consistency, we note that working at the level of second order regularity the condition (43a) relating the third to the second derivative of $F$ is assumed to hold. Differentiating (89) twice and evaluating the result at $r=\pi$, the third derivative of $F$ at the point arises. However, that third derivative term can be replaced via (43a) with a second derivative term, giving again an equation for $f''(\pi)$. The result is identical to (92). This is rather expected, given that at each given order of regularity, a number of odd derivatives of $F$ depend on lower even order derivatives at $r=\pi$, an information already encoded in the Karman-Howarth. Hence, relations involving those odd derivatives should not contain any new information.

For the third order regularity, the function $w(r)$ depends on the second and fourth derivative of $F$ at $r=\pi$, as can be seen by the equations (61). The first derivative of (79) at $r=\pi$ gives one relation for $f''(\pi)$ and $f^{(4)}(\pi)$. The third derivative of (89) at $r=\pi$ gives, using also the first order regularity condition (43a), a second relation for the quantities $f''(\pi)$ and $f^{(4)}(\pi)$. These two relations define $f^{(4)}(\pi)$ in a simple manner in terms of $f''(\pi)$. The quantity $f''(\pi)$ is defined now through a cubic equation. The result is explicit but it is rather too involved to quote here. In the application that follows it will become clear that we need not to use such a high order regularity; even for illustration purposes, applying regularity up to the second order makes all basic characteristics of renormalization manifest.

**D Application**

We have to solve equation (89) using the condition (81) of zero Loitsyansky integral, for different orders of regularity. For concreteness, we shall use the DNS data of [11][12] as a basis for input values. We consider the lowest Reynolds number case ($Re_\lambda \sim 100$) which allows us to go quite far into the dissipation range by only a few hundred wavenumbers. We set the viscosity at the value $v=2/1000$, and the kinetic energy at $K=5/10$. The dissipation $\varepsilon$ is to be found as part of the solution: the solution of the problem is a function $f(r)$ which solves equation (89) for the specific value of $\varepsilon$ for which $f(r)$ satisfies (81). As explained earlier, in section IV B, there is a correspondence between the initially set (and preserved by the forcing) value of $K$ and the final (steady state) value of $\varepsilon$. The only other numerical input required is the value of the closure parameter $\alpha$ introduced in equation (48) relating it to the inertial range Kolmogorov constant $C_2$. The constant $\alpha$ can also be expressed in terms of the skewness $S$ of the velocity-gradient distribution: $\alpha=(\sqrt{2}/9)(-S)$ as we discuss in Appendix D. For our purposes, we shall use the nominal value $-S=50/100$ rounding the quoted value in [12].

We start with the zeroth order of regularity. That is, we use the bare form of the closure for which $w(r)=r$. We need to find the value of $\varepsilon$ so that the associated solution $f(r)$ is consistent with the zero Loitsyansky integral condition (81). This can be done iteratively through a mid-point procedure. From the data of [11] the dissipation $\varepsilon$ is expected around 0.09. We start with a pair of values of $\varepsilon$ around 0.09 so that the Loitsyansky integral associated with each solution has a different sign, say 0.08 and 0.10, and calculate also the Loitsyansky integral associated with the solution for the mid-point, 0.09, of that interval. The half-interval whose end-points correspond to Loitsyansky integrals of different sign is picked as the new interval. We proceed this way until the Loitsyansky integral is small enough ($\sim 10^{-20}$ appears to be adequate for our purposes). This determines $\varepsilon$ also in some twenty significant figures.

We proceed to the first order of regularity. The function $w(r)$ is given by equation (54). The coefficient of $w(r)$ is a constant number and the method of solution is the same as with the zeroth order.

At the second order the coefficients $a_1$ and $a_2$ of $w(r)$ are given by (58), or equivalently in the present problem, by the simplified equations (94). They depend on the value of the field $f$ at $r=\pi$. The only modification we need to do here is to insert an additional set of iterations for each value of $\varepsilon$. For each value of $\varepsilon$ in the mid-point procedure, we start with $f_\pi=0$ in the expressions (94) and find associated solution $f(r)$ and the value of $f(\pi)$ for that solution. We set $f_\pi = f(\pi)$ in the expressions (94) and find a new solution. The new value of $f(\pi)$ provides a new value $f_\pi$ for the next step. We proceed iteratively until the input $f_\pi$ and the resulting $f(\pi)$ differ adequately little (an error of $10^{-25}$ arises after 20 iterations). This provides the second order regularity solution associated with the given value of $\varepsilon$. Applying the procedure at each step of the mid-point procedure for $\varepsilon$ we find the desired value of $\varepsilon$ and the associated solution. Thus the second order solution is characterized by the value of $\varepsilon$, same as the lower order solutions, as well as the value of $f_\pi$. The latter is required to specify the coefficients $a_1$ and $a_2$, which are functionals of the field $f$ at this order and are determined simultaneously with the solution. The obtained values for all these quantities are given in Table I, where we include the values corresponding to the lower orders, for completeness. The dissipation $\varepsilon$ is given in many decimal places which will be convenient below. The quoted decimal places of the results for the other quantities are indicative, serving only to illustrate their size and/or variation from order to order.

Table I. Solution parameters derived for each order of regularity.

| order | $\varepsilon$ | $a_1$ | $a_2$ | $f(\pi)$ |
|---|---|---|---|---|
| 0 | 0.08871786276 | 0 | 0 | −0.009818 |
| 1 | 0.08427980939 | $1/(3\pi^2)$ | 0 | −0.011373 |
| 2 | 0.08399199212 | $1.0764/(3\pi^2)$ | 0.00015687 | −0.011457 |

For each order of regularity, once the solution is obtained, the spectrum $E(k)$ is calculated by the formula (21). The spectra of energy, dissipation, palinstrophy and $k^6$-moment are given in Figure 3 for the zeroth, first and second order of regularity; the dissipation, palinstrophy and $k^6$-moment spectra are given in the dimensionless form (E1). The wavenumbers are quoted characteristically in the dimensionless form $k\eta$, where $\eta$ is the dissipation range Kolmogorov scale $\eta=(v^3/\varepsilon)^{1/4}$. We plot the absolute values of the spectra; beyond a certain wavenumber each $E(k)$ starts oscillating in the sign-changing fashion we encounter in section IV E3. The results realize fairly clearly the claims and conjectures we were lead to in section IV E3 through simple numerical and analytical examples. First, the spectra exhibit asymptotic irregularities in the form of oscillatory (sign-changing) power-laws, and second, those irregularities are gradually removed by applying higher order regularity conditions: The large $k$ behavior of the spectrum at the $n^{th}$ order of regularity is

$$E_{n-\text{order}}(k) = \text{constant}\,\frac{(-1)^{[k]}}{k^{2n+2}} \qquad (95)$$

This becomes clear in the Figures 3b to 3d, where the graph of the spectrum of the $k^{2(n+1)}$-moment, that is, the quantity $k^{2n+2}E(k)$, becomes a straight horizontal line in the large $k$, for $n=0,1,2$. This fact is also verified by the numbers: in the regimes shown, the absolute value of the spectra $k^{2n+2}E(k)$ is constant in some six significant figures.

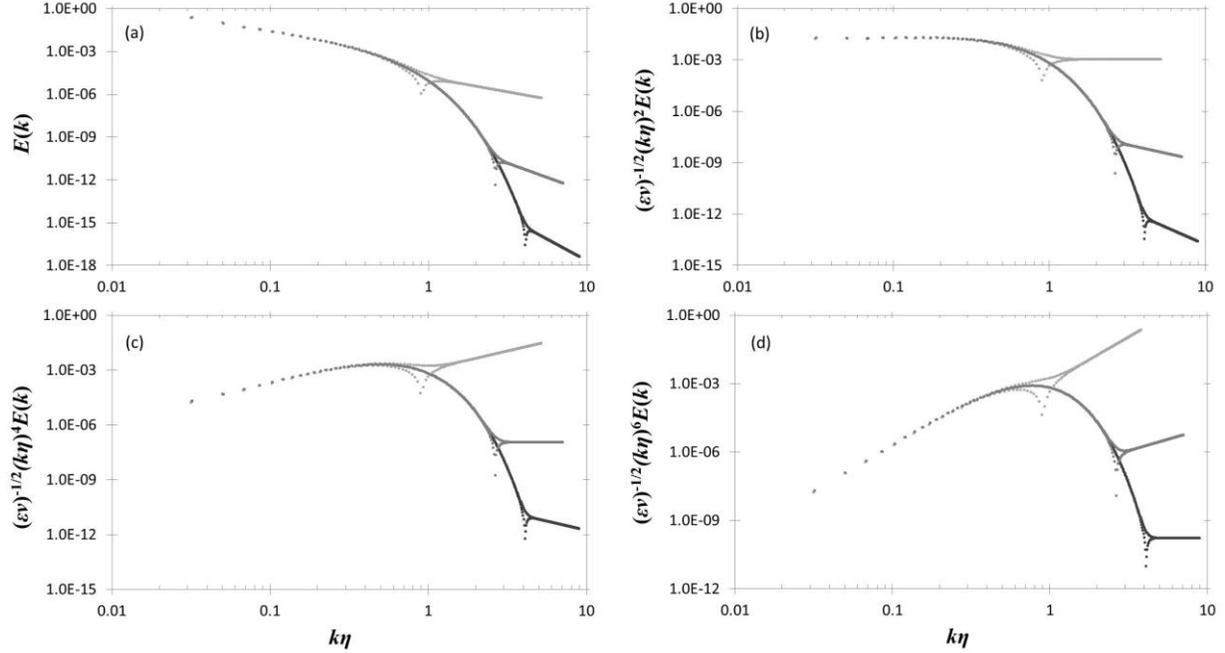

Figure 3. Plots of the absolute value of the spectrum $E(k)$, dimensionless dissipation spectrum $(\varepsilon\nu)^{-1/2}(k\eta)^2 E(k)$, dimensionless palinstrophy spectrum $(\varepsilon\nu)^{-1/2}(k\eta)^4 E(k)$ and dimensionless $k^6$-moment spectrum $(\varepsilon\nu)^{-1/2}(k\eta)^6 E(k)$ against $k\eta$, are shown in the graphs (a) to (d) respectively, for the zeroth, first and second order of regularity, indicated by different tones of grey dotted lines. The order increases downwards, showing the advancement of the physical regime deeper in the dissipation subrange. The softening of the irregularities in the large $k$ regime, formalized by (65), is also visible.

Recalling also the discussion of section IV E4 we observe that the $k^{2i}$-moment is well defined in the strong sense for all $i<n+1$. Also, in accordance with our reasoning there, increasing the order of regularity the regime of wavenumbers where the spectrum remains physical (i.e. positive definite and decreasing for large $k$) is extended deeper in the dissipation subrange. Indeed, this is clearly visible in the graphs of Figure 3, and we find specifically that at the zeroth order of regularity the spectrum is physical up to $k\eta=0.35$, in the first order up to $k\eta=2.2$ and at the second order up to $k\eta=3.7$. This is a most important result, exhibiting the effect of 'renormalization', as we have called it. An empirically postulated closure, such (48), is 'unaware' of the subtleties of the compact space, leading to irregular behavior of the spectrum in the dissipation subrange. The subtleties are a series of constraints imposed by the compact space on the fields $F$ and $H$, involving increasingly higher derivatives of the fields, given by (42). The problem is also the key to its solution: adding suitable corrective terms to the original closure any specific number of such constraints can be satisfied. That turns the spectrum physical up to a large wavenumber, which increases with the number of satisfied constraints. These results also illustrate the following idea: there is nothing wrong with a spectrum becoming unphysical in the large wavenumbers, as long as the asymptotic behavior does not lead to greater consistencies apart from the spectrum being essentially undefined beyond a certain wavenumber and, also, as long as that wavenumber can be pushed as deep in the dissipation subrange as we want. The point is that, from practical point of view, we are interested in finite wavenumbers, and from theoretical point of view this kind of spectra are still manageable. One could put forward a working definition that such spectra are the ones whose all energy moments are at least weakly defined.

We now turn to the consistent calculation of those moments and – through them – the quantification of the information carried by the spectrum in its physical regime. For each order of regularity, the results for the first three energy moments i.e., the dissipation $\varepsilon$, the (dimensionless) palinstrophy and the (dimensionless) $k^6$-moment are given in Table II. The palinstrophy and the $k^6$-moment in dimensionless form are defined by the left hand side of equation (E1) in the Appendix, for $i=2,3$, and they are calculated by the right hand side of the relations (E2) and (E3) respectively, once the value of the dissipation $\varepsilon$ has been found. The results are quoted in a relatively large number of decimals; in fact, given that $\varepsilon$ has been determined by even more decimals, the moments can be determined through the right hand side of (E2) and (E3) with even higher accuracy. The number of the significant figures we choose to quote is explained below. The spectral (i.e., left hand) side of equalities (E2) and (E3) are a specific realization of equation (75) and must be understood in a regularized sense, as implied by (75). In practice, where we use finite sums, regularization is implemented very effectively by the re-summation formulas of Appendix F. Equalities (E2) and (E3), as well as the relations

$$\sum_k E(k) = K, \qquad 2\nu \sum_k k^2 E(k) = \varepsilon, \qquad (96)$$

are verified in the quoted decimals by including wavenumbers up to $k\eta \sim 5$ (the total energy is calculated even more very accurately, thus we give it as an exact fraction value). In the second order of regularity, the simplest re-summation [$m=1$ case of formulas (F2) or (F3)] is adequate to obtain the $k^6$-moment in the quoted decimals, while the dissipation and the palinstrophy relations are verified in even higher accuracy. In the first order of regularity, the dissipation is obtained by the $m=1$ of formulas (F2) or (F3) essentially in the quoted decimals, while the higher moments require the $m=2$ re-summation formulas. The zeroth order requires consistently higher $m$ in order to verify the relations in the given decimals. In, all, using a suitable regularization, one may check and verify to high accuracy the consistency of the solution in $k$- and $r$-space.

In Table III we give the results for the moments calculated from the physical regime alone. We quote the result in the number of significant figures for which they agree with the accurate values (in the given decimals) of the moments shown in Table II. The zeroth order results from the physical regime are quite poor, due to the fact that physical regime at that order does not even reach the peak of the palinstrophy spectrum. Specifically, the total energy and the dissipation turn out to be 0.49 and 0.057 respectively. For the first and second order, one observes the high accuracy with which the moments are obtained from the physical regime of the spectrum. Clearly, this is possible due to depth of the physical regime already in the first order. The number of figures we use in Table II is a merely uniform choice which is required for quoting the correct significant figures of the most accurate result of Table III; after all, the moments have been uniformly determined by almost twice as many decimals for all these cases. One should also note the rather significant level of accuracy by which palinstrophy is determined in the first order of regularity in the physical regime. We consider that an implication of the mild divergence of the palinstrophy spectrum at that order: $k^4 E(k)$ approaches $(-1)^{[k]}$ for large $k$, therefore palinstrophy is what we called in section IV E4 marginally ill-defined (in the strong sense). The same can be noted for the value of the $k^6$-moment in the second order of regularity. On the other hand, one may note that in the first order of regularity, the spectrum $k^6 E(k)$ diverges as $(-1)^{[k]} k^2$ which is a strong divergence. The implication is that the associated moment, although is represented quite decently in the physical regime by three correct significant figures, it is quite less accurately represented than palinstrophy.

Table II. Energy moment values derived for each order of regularity.

| order | $K$ | $\varepsilon$ | palinstrophy | $k^6$-moment |
|---|---|---|---|---|
| 0 | 5/10 | 0.08871786276 | 0.07581334077 | 0.03670348650 |
| 1 | 5/10 | 0.08427980939 | 0.07582647508 | 0.03673808220 |
| 2 | 5/10 | 0.08399199212 | 0.07582736264 | 0.03674072099 |

Table III. Physical regime values of the energy moments.

| order | $K$ | $\varepsilon$ | palinstrophy | $k^6$-moment |
|---|---|---|---|---|
| 0 | 0.5 | 0.1 | | |
| 1 | 0.50000000 | 0.0842795 | 0.075814 | 0.0367 |
| 2 | 0.500000000000 | 0.0839919921 | 0.07582736 | 0.0367407 |

We may note at this point that the value $\varepsilon \sim 0.09$ we find as part of the solution, is consistent with the DNS value for the $Re_\lambda \sim 100$ case of [11] which we used as a reference. That in turn means that other global characteristics, that is, the Reynolds number $Re_\lambda$ and the dissipation length scale $L_\varepsilon = (2K)^{3/2}/\varepsilon = 1/\varepsilon$, which are specified by $\varepsilon$, are also consistent with [11]. The values of these quantities for the different orders of regularity are given in Table IV. Therefore we understand that the toy model (48), although it does not encapsulate finer phenomena of turbulence such as the bottleneck effect or anomalous scaling [4][2], it does capture well the relation between $K$ and $\varepsilon$ and hence the global characteristics of the flow. We may also note that other more intrinsic characteristics are also reproduced: the peaks of dissipation and palinstrophy spectrum arise roughly at $k\eta \sim 0.2$ and $k\eta \sim 0.5$ respectively, consistently with the behavior of the DNS flows [11][12]. Hence, the model (48) can be the basis for more elaborate models as we discuss in the next section. In Table IV are also included the values of the integral scale $L_{11}$ for each order of regularity. The integral scale is naturally defined here as the integral of $f(r)$ from $r=0$ to $r=\pi$. In the Appendix G we re-write this definition also in spectral space, deriving a formula quite similar to the standard spectral formula for $L_{11}$ in isotropic turbulence. The data of [11] show that $L_{11} \sim 1.1$ in the DNS, in contrast to $L_{11} \sim 0.9$ obtained here. The integral scale is very strongly affected by the very first few wavenumbers. The discrepancy may be attributed essentially to the existence of the $k=1$ modes in the DNS, as we discussed in section V A. That could be corrected by organizing and labeling differently the shells, especially the first one, of the DNS. Hence, we regard the discrepancy more as a matter of definition than an essential one. At any rate, one should bear in mind that the integral scale is strongly affected by the most anisotropic modes of the DNS and its value is by nature an imprecise characteristic of the flow.

The results given in Table III can be read off from another point of view, that is, as illustrating the degree of *convergence* of the results according to the *order of regularity*. One observes that the change of these quantities, which are essentially determined through the dissipation, is in the order of one part in a thousand, as we go from the first to the second order of regularity. The dissipation, being the lowest energy moment, is affected the most by renormalization, which changes slightly the spectrum all the way to the smallest wavenumbers. As we see in Table III, palinstrophy changes in the order of one part in ten thousand, from the first and second order of regularity. The convergence of the moments quantifies the convergence of spectrum towards a limiting spectrum which remains physical for all wavenumbers. The latter statement is clearly an important assumption underlying everything we do here, but it also the most difficult to handle mathematically. The fact that the spectrum does not change significantly as the physical regime is extended in larger wavenumbers by regularity is already visible in the graphs of Figure 3. The moments give a more specific measure of convergence. One may note the relatively slow convergence of the $k^6$-moment at these orders. That can be regarded as another implication of the fact that in the first order of regularity the spectrum $k^6 E(k)$ diverges strongly, as we noted above, discussing the relatively low accuracy of the representation of this moment in the physical regime of the spectrum. In all, we may say, taking now explicitly into account convergence, that values of the dissipation, dimensionless palinstrophy and dimensionless $k^6$-moment are 0.0840(2), 0.07583(1) and 0.0367(1) respectively, with an estimated error in the last digit given in the brackets.

Table IV. The Reynolds number, the dissipation length scale and the integral scale derived for each order of regularity.

| order | $\varepsilon$ | $\text{Re}_\lambda$ | $L_\varepsilon = (2K)^{3/2}/\varepsilon$ | $L_{11}$ |
|---|---|---|---|---|
| 0 | 0.08872 | 96.92 | 11.27 | 0.84724 |
| 1 | 0.08428 | 99.44 | 11.87 | 0.87392 |
| 2 | 0.08399 | 99.61 | 11.91 | 0.87557 |

Regarding the energy transfer $T(k)$, the energy balance equation (78) implies that at the steady state we have

$$T(k_1) = 2\nu k_1^2 E(k_1) - \varepsilon, \text{ for } k = k_1, \qquad T(k) = 2\nu k^2 E(k), \text{ for } k > k_1 \qquad (97)$$

which encodes the conservation of energy by the cascade processes

$$\sum_k T(k) = 0 \qquad (98)$$

and also verifies the relation (68) between energy spectrum and energy transfer for large $k$ that came through our discussion in section IV E3. At the steady state, the energy transfer functions differs by a factor of $k^2$ from the energy spectrum, for the wavenumbers which are not forced or, in general, if one goes to large enough wavenumbers, as it is implied by the asymptotic formulas we have derived here in various ways.

Finally, the effect of the Reynolds number on renormalization can be understood in the following way. The coefficients of the function $w(r)$ of the renormalized cascade are of order one, in units of the size of the compact space. Introducing a dissipation scale dimensionless distance $x$ by $r=\eta x$, we have that $w=\eta x(1-O(\text{Re}_\lambda^{-3})x^2+O(\text{Re}_\lambda^{-6})x^4+\ldots)$. We may then say that the counter-terms are increasingly lighter regarding the large wavenumber effects for higher Reynolds numbers, that is, any given counter-term is increasingly less necessary for removing the irregularities from the spectrum for any given wavenumber $k\eta$. Put differently, the physical regime associated with any given order of regularity, becomes larger and larger as we increase the Reynolds number. Indeed, for the first order of regularity we mentioned above that for $\text{Re}_\lambda \sim 100$ the physical regime extends to the wavenumber $k\eta=2.2$. Using again data from [11][12] as sensible input, we find that for $\text{Re}_\lambda \sim 200$ the physical regime the physical regime extends to $k\eta \sim 2.5$ and for $\text{Re}_\lambda \sim 400$ to $k\eta \sim 2.8$, for the same order of regularity. These results tie well with the interpretation of our renormalization as a perturbative procedure with the inverse Reynolds number as the small parameter (section IV E3).

On the other hand, the rescaled distance argument we used above should be carefully applied. In compact space, the effects of large and small distances are not cleanly separated, as it is very much the case in infinite space. We have experienced this fact throughout this work, where the – figuratively – large distance differentiability problems of the field $F$ at $r=\pi$, affect the *large k* behavior of the spectrum. Hence the naïve conclusion deriving from the above argument, that for very large Reynolds numbers one may not need even the first regularity condition, is wrong, as the regularity conditions that operate through $r=\pi$, are inescapable. In fact, for the truth of the matter, one finds that e.g. for $\text{Re}_\lambda \sim 400$ the first negative values of the spectrum arise at somewhat larger wavenumbers, $k\eta \sim 1.1$ as opposed to $k\eta \sim 0.94$ of the $\text{Re}_\lambda \sim 100$ case, but the uniform monotonicity of the spectrum is ruined in the strict sense due to small amplitude oscillations which start already from the smallest wavenumbers. Hence, strictly speaking, the spectrum is not physical anywhere at the zeroth order of regularity for Reynolds numbers in the range of hundreds or higher. Presumably, due to such phenomena, we have refrained from giving a too rigorous definition of what is physical, because, after all, whether or not small oscillations can be tolerated in a *positive-definite* spectrum is a matter of realistic assessment. Also, this type of irregularities in the regime of small wavenumbers, e.g. on the left of bottleneck effect peak, are very much present in the DNS spectra.

# VII DISCUSSION

The original motivation for this work was provided by the following problem. Consider DNS flows governed by the incompressible Navier-Stokes equations (1) where we have added on the right hand side a forcing term of the form $\Gamma u_i$, where $\Gamma$ is a constant with dimensions of inverse time. This system leads to somewhat strongly oscillating steady state [13]. Consider neglecting the DNS boundary conditions (8), which spoils isotropy and derive the corresponding Karman-Howarth equation. One obtains equation (4) with a forcing term $2\Gamma F(r)$ on the right hand side, while the energy balance equations (1) and (23) acquire terms of the form $2\Gamma K$ and $2\Gamma E(k)$ on their right sides. The spectral energy balance is similar to (78), with a forcing rate of energy fed into the system of the form $\varepsilon_k = 2\Gamma E(k)$, only $k$ is the usual continuous wavenumber of isotropic turbulence. That is, there is uniform feeding of energy at all scales in this 'linear forcing' scheme, introduced in [13]. In the framework of the Karman-Howarth equation, which lives in infinite space, the problem with this forcing is that it does not possess a fixed length to provide the scale of the integral scale of turbulence, as opposed to the DNS flows which have one by construction. It is by no means obvious that there is a steady state for this system in the usual isotropic framework; in fact, most likely there is not, given the lack of the fixed large length scale in the system. Imagine now we force the existence of a steady state, the idea being that we let the system to instruct us what kind of boundary conditions, or better, asymptotic behavior, would require so that to possess a steady state. Clearly those boundary conditions should first of all introduce a large length scale.

Indeed, using the closure (48) one can write down the steady state Karman-Howarth equation for this system, which is a second order ordinary differential equation for $F(r)$. The first thing one can do is to look at the large distances behavior of the solution. That can be done analytically. For large distances $F(r)$ is small compared to $F(0)=u'^2$ therefore the formula (48) simplifies to the linear relation $H=-(3/2)\alpha u'rF'(r)$. Also the associated term i.e., the cascade term, dominates over the dissipation term in the Karman-Howarth equation. Thus, cascade is balanced by the forcing term $2\Gamma F(r)$ in the large distances. The result is a linear equation whose solution reads $F(r) \sim r^{-2}J_4(\sqrt{k_0}r)$, where $J_4$ is the Bessel function of the first kind and $k_0$ is a constant which depends on $\alpha$, $u'$ and $\Gamma$. This function is oscillatory (with decreasing amplitude), passing through zero infinitely many times as $r$ goes to infinity. The first zero occurs at a distance of the order of $k_0^{-1}$. One realizes that a length scale introduces itself with this behavior. As we discussed in section V B, precisely this kind of behavior arises by introducing a fixed length scale in infinite space through a small wavenumber cutoff in the spectrum. Of course, as we mentioned in section V B, such configurations are unstable due to backscattering. Presumably, the small wavenumber spectrum of the function $F(r) \sim r^{-2}J_4(\sqrt{k_0}r)$ possesses unphysical terms such as $k \cos(k_0/k)$. In all, one concludes that the cutoff must be permanent, and the most natural way to do that is by turning the spectral space discrete through boundary conditions on $F(r)$ at some explicit finite distance. Regarding the Karman-Howarth equation as an essentially second order equation for $F(r)$, periodicity induces such boundary conditions equally naturally, in some sense, cutting off such asymptotically oscillatory functions at their first minimum i.e., at the minimum beyond their first zero. The periodicity of $F(r)$ is explicitly expressed by (9a) which, by the evenness of this function, implies (9b); the triple correlation function $H(r)$ enters the Karman-Howarth equation through its first derivative and needs only be periodic in the sense implied by equations (9). Periodic boundary conditions reflect the DNS flows and, as discussed in section III, they are a most convenient way to turn an infinite space to a finite one: *compactification* preserves the closedness of the infinite system, as it introduces finiteness without actually introducing boundary. Thus, periodicity, imposed at the level of the correlation functions, becomes the basis for this work.

Once this is decided, and if one is interested in following this line of thought, everything else follows almost inevitably. Central to our reasoning are the eigenfunctions $f_k(r)$, given by (11), of the free (non-interacting part), that is, the pure dissipation part of the Karman-Howarth equation. They essentially contain all the useful information: the spectral space, given by (14); the orthogonality of the modes and the decoupling of the mode $k=0$, which induces the vanishing of the compact space Loitsyansky integral, as discussed in section IV B; the fact that $F(r)$ is periodic only as $C^2$ differentiability class function (and $H(r)$ only $C^1$); the 'obstruction' to smooth periodicity relations (37) which induce the regularity relations for $F(r)$ and $H(r)$ given in (42) i.e., the necessary conditions for the spectrum and energy spectrum to be physically realizable, as discussed in section IV D. With these results and tools at hand one can build the rest of theory.

The first thing one observes is that an $F(r)$ which would produce a perfectly physical spectrum in infinity space, produces an asymptotically oscillatory spectrum of the form $(\pm)/k^2$ in compact space, where the irregular/unphysical behavior starts within the dissipation subrange. The effect of the regularity conditions is, (i) to soften that behavior, so that if $n$ conditions are satisfied then the dominant large wavenumber behavior of the spectrum is $(\pm)/k^{2n+2}$ and, most importantly, (ii) to push the first appearance of unphysical behavior deeper in the dissipation subrange, rendering the rest of the spectrum regular, that is, positive and decreasing in the dissipation range. A fairly important clarification is that the softening of irregularities allows more energy moments to be defined in the usual strong sense, but all energy moments are defined in a weak (regularized) sense, thereby allowing the asymptotically oscillatory spectra to be consistent with the analyticity of $F(r)$ in the neighborhood of $r=0$. Analyticity around $r=0$ is a property of $F(r)$ in ideal isotropic turbulence which one would like to preserve, and is also an implication of the Karman-Howarth equation under a reasonable closure scheme. The fact that more moments are defined in the strong sense is consistent with the fact that the physical regime of the spectrum is extended deeper in the dissipation range as we increase the order of regularity (that is, the number of regularity conditions satisfied). Similarly, the energy transfer function behaves as asymptotically $(\pm)/k^{2n}$ for order $n$ regularity imposed on $H(r)$, rendered regular for increasing larger $k$ as we increase the order $n$. These observations are then used to 'round' the closure scheme so that to conform to the compact space subtleties. A closure scheme, which is a functional relationship $H=H[F]$, does not a priori respect regularity, but it can be made to respect it at any desired order by adding corrective terms, constructed systematically by imposing that regularity condition on $F$ and $H$ at that order. This is the process we call as renormalization, borrowing a term from quantum field theory. Section IV E is devoted to developing these ideas in an explicit form.

These ideas must be tested in a situation where the actual (numerical) solution of the Karman-Howarth equation exhibits the theoretically expected behavior. The DNS flows, with their usual low-wavenumber forcing, clearly provide a most characteristic application. This situation provides also a good opportunity to show how DNS flows of interest are – finally – transcribed into our language. This is done in section VI. In the same section the theoretical expectations on renormalization are verified in great detail, regarding both the effect of regularity on the physical regime of the spectrum as well as the well-defined-ness of, or better, the accuracy of knowing, the energy moments for a given order of regularity.

Attempting to think our construction through a single picture one may note the following. Compactifying the theory, given by the Karman-Howarth equation as the dynamical equation for the field $F$, one stumbles upon the obstacle of differentiability: $F$ can only be twice differentiable as a periodic function, because the third and all the higher odd derivatives of $F$ do not vanish at the end-points $r=\pm\pi$, as it be would the case of an even function of period $2\pi$ which is infinitely smooth, e.g. such as $\cos(r)$. If that were the last word one could say, then the whole notion of periodicity would be somewhat incomplete. One can differentiate the Karman-Howarth equation and relate different order derivatives of $F$, and if higher odd derivatives are simply left in fate, it feels that one does not know how to actually 'sow' the theory at the points $r=+\pi$ and $r=-\pi$. There must be some rules regarding those derivatives, dictated by the equation itself, which $F$ should satisfy at all times. It makes sense that such rules should be respected by the free Karman-Howarth, which also preserves them through time due its linearity. The mode relations (37) encode precisely that information. When combined with the condition of regularity of the spectrum (and energy transfer) one obtains the sought for rules given by (42): although the higher odd derivatives of $F$ (and $H$) are not continuous – and therefore not zero – across $r=\pm\pi$, their discontinuities are constrained by a sequence of relations, the first three of which are given in (43) and (44). It is interesting to note that it is the a priori necessary physical condition of regularity/realizability that completes the compactification. Although the point of view expressed by these observations may sound subjective and more of a matter of taste, its operational importance is realized immediately in the presence of interactions. A generic closure, however reasonable might be empirically, messes up the spectrum leading to the irregular behavior of asymptotic oscillations. This is handled by merely applying consistently the requirements of compactification on the both the fields $F$ and $H$, realized through the process of renormalization of the closure. Along the way, an interesting idea is introduced, that there is nothing wrong with allowing unphysical behavior of the spectrum in the infinitely large $k$, as long as we have physical behavior in a realistically useful regime i.e., a few $k\eta$, whose depth we can control, and the whole thing makes sense mathematically for all $k$.

The artificial part of the whole construction clearly is the closure scheme, given here by (48). There is nothing inevitable about it, neither empirically complete, as it does not accommodate the bottleneck effect and anomalous scaling phenomena observed in turbulent flows [4][2]. It possesses the convenient characteristic that it involves only the first derivative of $F$, not spoiling the second order nature of the free Karman-Howarth equation. Closures involving higher order derivatives of *F* are perfectly meaningful and our methods can certainly be applied, but they would require a separate analysis regarding their behavior under renormalization. Thus, we will leave them aside for future work. Let us then focus on closures on the same family as the toy closure (48). Such closures can be written as

$$H = G_1(r) G_2(F(0) - F) F'  \qquad (99)$$

where $G_1(r)$ is a function of *r* with coefficients which in general depend on the Reynolds number, and $G_2$ may be a simple function with specific scaling properties. The time dependence of all quantities is left understood, as usually in this work. Closures such as (99) are general enough to accommodate the bottleneck and anomalous scaling phenomena. Also one needs greater freedom, than it is available in (48), in order to encode properly the behavior of the energy spectrum of actual flows in large wavenumbers. For example, the relation (D3) between skewness and the Kolmogorov constant is far too restrictive. Such problems are overcome easily by generalizations such as (99). Renormalization can be applied on (99) without much greater difficulty than working with (48). We may introduce the renormalized, or 'dressed', form of the closure as

$$H = G_1(w(r)) G_2(F(0) - F) F'  \qquad (100)$$

where the function $w(r)$ must be determined by the regularity conditions. It is easy to show that the first order regularity is satisfied if $w'(\pi)=0$, that is, it is satisfied by the same simple form of $w(r)$ which we used for the toy closure (48), given by equation (55). This holds for any useful form of the functions $G_1$ and $G_2$. At the second order of regularity the determination of the coefficients $a_1$ and $a_2$ of the function $w(r)$ cannot be obtained in an explicit form as in (58) and the respective conditions must be solved numerically. Nonetheless, already the first order regularity goes a long way producing practically useful results, as long as comparison with results from direct numerical simulations is concerned. Presumably, one should note that, if we restrict ourselves to the first order, different forms of the function $w(r)$ which satisfy $w'(\pi)=0$ could possibly be more effective than the polynomial form given in (55). For example, one may investigate the effectiveness of a function such as $2\sin(r/2)$. Also, from an entirely different point of view, once one knows how to handle and express the compactified theory in both the *r*- and *k*-space, one could very well choose to postulate closures directly in spectral space, defined over the wavenumbers introduced in this work. Such matters shall be left for future work.

## APPENDIX

## A  Properties of the wavenumbers *k*

The wavenumbers are defined by the boundary condition $f_k'(\pi) = 0$ where the eigenfunctions $f_k(r)$ are defined by (11). The eigenfunctions can be written as

$$f_k(r) = \frac{2}{kr} j_1(kr) \qquad (A1)$$

where $j_n(x)$ is the order *n* spherical Bessel function of the first kind; the definition and properties of the Bessel functions can be found e.g. in [8]. The Bessel functions of different orders are related by recursive relations; here we shall need the fact that

$$\frac{1}{x}\frac{d}{dx}\left(\frac{j_1(x)}{x}\right) = -\frac{1}{x^2} j_2(x) \qquad (A2)$$

Then (A1) and (A2) imply that the wavenumbers are defined by the relation $j_2(k\pi)=0$. On the other hand, the spherical Bessel are related to the Bessel functions of the first order by

$$j_n(x) = \sqrt{\frac{2}{\pi x}} J_{n+\frac{1}{2}}(x) \qquad (A3)$$

which means that *k* are defined by

$$J_{5/2}(k\pi) = 0 \qquad (A4)$$

The virtue of relating *k* to the roots of a known function is that they become in some sense known objects. For example, *Mathematica* can give us the *k*'s to a great many decimals by the formula BesselJZero[5/2,*i*]/$\pi$ for *i*=1,2,3,…

Explicitly, the wavenumbers *k* are defined by equation (14), which can also be re-written in the form of (40). Squaring (40) we may solve for sin($k\pi$) to obtain

$$|\sin k\pi| = \frac{1}{\sqrt{1 + \left(\frac{1}{k\pi} - \frac{k\pi}{3}\right)^2}} \qquad (A5)$$

This equation shows that for large $k$ the $\sin(k\pi)\to 0$, that is, $k$ approaches an integer. The numerical values show that $k$ is always something less than an integer, from which we understand the sign of $\sin(k\pi)$:

$$\sin k\pi = \frac{(-1)^{[k]}}{\sqrt{1+\left(\frac{1}{k\pi}-\frac{k\pi}{3}\right)^2}} \tag{A6}$$

where $[k]$ and the integer part of $k$ which acts as a counting index for the $k$'s: $[k]$=1,2,3,… From (A2) we have that for $k\to\infty$

$$(-1)^{[k]} k\pi \sin k\pi \to 3 \tag{A7}$$

a relation which is very useful for large $k$ asymptotic considerations. For large values of $k$, we may write $k=n+\delta k$, where $\delta k$ is small compared to the integer $n$. If we insert this expression into (14) or (40), or even more simply, into (A7), and expand for small $\delta k$, it is easy to show that $\delta k=-3/(n\pi^2)$, obtaining the first correction to the leading behavior of $k$ which we have already given in equation (16). Note that $[k]=n-1$.

**B  Orthogonality relations**

Let $k$ and $q$ two arbitrary wavenumbers of the compact space. Recall the eigenvalue equation (12)

$$[r^4 f'_k]' + k^2 r^4 f_k = 0, \qquad [r^4 f'_q]' + q^2 r^4 f_q = 0 \tag{B1}$$

which we re-write using primes to denote differentiation with respect to $r$. We multiply each equation of (B1) with the eigenfunction of the other, and re-write their first term:

$$[r^4 f'_k f_q]' - r^4 f'_k f'_q + k^2 r^4 f_k f_q = 0, \qquad [r^4 f'_q f_k]' - r^4 f'_q f'_k + q^2 r^4 f_q f_k = 0, \tag{B2}$$

We integrate each equation from $r=-\pi$ to $r=\pi$. By the boundary condition (13), and the evenness of the eigenfunctions, the first terms of these equations vanish. The second terms are identical, that is, they go away by subtracting the resulting relations, and we obtain

$$(k^2 - q^2)\int_{-\pi}^{\pi} r^4 f_k f_q \, dr = 0 \tag{B3}$$

which means that the eigenfunctions corresponding to different wavenumbers are orthogonal in the sense of (B3). When the two wavenumbers are the same we obtain the normalization of the associated eigenfunction, given by (19). The final formula for the coefficient $c(k)$ in (19) is obtained by a little algebra and using also the equation (40).

Diving each of the equations (B2) with the respective wavenumber squared and following the same steps we obtain

$$-(k^{-2} - q^{-2})\int_{-\pi}^{\pi} r^4 f_k' f_q' dr = 0 \tag{B4}$$

only this time the result comes from the middle term of each equation in (B2). Thus we obtain that also the derivatives of the eigenfunctions are orthogonal. Their normalization is derived by setting the two wavenumbers in one of the equations (B2) equal and integrating from $r=-\pi$ to $r=\pi$:

$$\int_{-\pi}^{\pi} r^4 f_k' f_k' dr = k^2 \int_{-\pi}^{\pi} r^4 f_k f_k dr \tag{B5}$$

that is, it is similar to (19) times a factor $k^2$.

## C Loitsyansky integral and eigenfunctions

In the main text, section IV B, we used the orthogonality relation so show that eigenfunctions with non-zero $k$ have zero Loitsyansky integral, expressed by equation (20). We would like to revisit this matter and discuss it from the point of view of the eigenvalue equation. Recall the eigenvalue equation (12)

$$[r^4 f_k']' + k^2 r^4 f_k = 0 \tag{C1}$$

We integrate both sides of (C1) from $r=0$ to $r=r$. We have

$$r^4 f_k'(r) - 0 + k^2 \int_0^r r^4 f_k(r) dr = 0 \tag{C2}$$

bearing in mind that the eigenfunctions are analytic functions and therefore their derivatives are finite, in fact $f_k'(0) = 0$, as the eigenfunctions are even.

Now at $r=\pi$ the first term vanishes by the boundary condition (13). Then the second term gives us equation (20), i.e., the compact space Loitsyansky integral of every non-zero mode vanishes, bearing in mind that the evenness of the modes allows us to work with the half interval as well. As it is discussed in detail in section IV B this means that all the dynamical degrees of freedom contributing in the field $F$ have zero Loitsyansky integral, which is very descriptively expressed through their orthogonality to the decoupled zero mode.

## D Skewness factor and the Oberlack-Peters model

The small $r$ behavior of the second and third order structure functions is given by

$$\langle (u_l(\vec{r})-u_l(0))^2 \rangle = \frac{\varepsilon}{15\nu} r^2 + \cdots, \qquad \langle (u_l(\vec{r})-u_l(0))^3 \rangle = S\left(\frac{\varepsilon}{15\nu}\right)^{3/2} r^3 + \cdots \qquad (D1)$$

where the first relation follows from the first equation in (47) and equation (6), while the second follows from the first and the definition of the skewness $S$ of the velocity-gradient distribution

$$S = \lim_{r\to 0} \frac{\langle (u_l(\vec{r})-u_l(0))^3 \rangle}{\langle (u_l(\vec{r})-u_l(0))^2 \rangle^{3/2}} \qquad (D2)$$

Inserting (D1) into the closure scheme (48) we obtain

$$\alpha = \frac{\sqrt{2}}{9}(-S), \qquad S = -\frac{12}{5C_2^{3/2}} \qquad (D3)$$

This is an equivalent but more descriptive expression for the model parameter $\alpha$. The Kolmogorov constant $C_2$ is a quantity meaningful at virtually infinite Reynolds numbers, and its meaning is less clear for Reynolds numbers which usually appear in the current DNS. One the other hand, skewness is a very specific characteristic of the (DNS) spectrum, and a standard output of the DNS, which is also related directly to the palinstrophy of the spectrum by the relation (E2) below.

## E Steady state formulas for the higher energy moments

The energy moments, defined by (32), are related to the derivatives of the field $F$ at $r=0$, the relationship following from equation (34). The first few explicitly relations are given in (36). Therefore differentiating the steady state Karman-Howarth equation (85) at the desired order and evaluating the result at $r=0$ we may express any specific moment in terms of the input and steady state parameters. It convenient to quote the result in dimensionless form, or more specifically, to scale the moments using the dissipation $\varepsilon$ and the viscosity $\nu$. Using the dissipation range Kolmogorov scale $\eta=(\nu^3/\varepsilon)^{1/4}$ we define the dimensionless moments by

$$\frac{\eta^{2i}}{\sqrt{\varepsilon\nu}} \sum_k k^{2i} E(k) \qquad (E1)$$

One should bear in mind that this formula should be understood in some regularized sense, as in formula (75). Then, indeed, by differentiating suitably (85) and using the formulas in (36) we obtain

$$\frac{\eta^4}{\sqrt{\varepsilon\nu}}\sum_k k^4 E(k) = \frac{7}{4}\sqrt{\frac{3}{10}}\alpha + \frac{k_1^2 \eta^2}{2} \tag{E2}$$

for the palinstrophy, and

$$\frac{\eta^6}{\sqrt{\varepsilon\nu}}\sum_k k^6 E(k) = \frac{189\alpha^2}{32} + 63\sqrt{\frac{3}{10}}\alpha a_1 \eta^2 + \frac{9}{8}\sqrt{\frac{15}{2}}\alpha k_1^2 \eta^2 + \frac{k_1^4 \eta^4}{2} \tag{E3}$$

for the $k^6$-moment of the spectrum. The closure scheme parameter $\alpha$ is related to the skewness factor $S$ by the relation (D3). The coefficient $a_1$ is the first counter-term coefficient i.e., the coefficient of the $r^3$ term of the function $w(r)$ for any given non-zero order of regularity [for the first two orders of regularity, $w(r)$ is given by equations (55) and (57)]. The terms involving $k_1$ come from the forcing term of the Karman-Howarth equation and are practically negligible, as they are negative powers of the Reynolds number.

## F Regularization of sums

In actual calculations the spectrum is known for a finite number of wavenumbers say, $k_1,\ldots,k_N$. Then the moments are defined by the partial sums

$$s_N = \sum_{I=1}^{N} k_I^{2i} E(k_I) \tag{F1}$$

Due to the oscillatory nature of the spectrum for large wavenumbers these sums also oscillate, too little or too much, depending on whether the moment is well defined in the strong or weak sense, as we explained in section IV E4. When the moment is defined in the weak sense, a very effective regularization is a weighted re-summation of its series:

$$2^{-m}\sum_{n=0}^{m}\binom{m}{n} s_{N-n} \tag{F2}$$

Re-summation methods are used in the theory of divergent series [14] as a means to give a meaningful definition to such series. Also, applied in convergent series they expedite their convergence. For $m=1,2,3,\ldots$ (F2) reads

$$\tfrac{1}{2}(s_{N-1}+s_N), \quad \tfrac{1}{4}(s_{N-2}+2s_{N-1}+s_N), \quad \tfrac{1}{8}(s_{N-3}+3s_{N-2}+3s_{N-1}+s_N), \quad \ldots \tag{F3}$$

The idea is that a sign-changing divergence can be trimmed though such combinations without affecting the convergent positive-definite part of the result. For our purposes the simplest ($m=1$) combination is mostly adequate.

**G Integral scale**

The integral scale in compact space is given by

$$L_{11} = \int_0^\pi f(r)dr = \int_0^\pi \frac{F(r)}{F(0)}dr = \frac{2}{3u'^2}\sum_k E(k)\int_0^\pi f_k(r)dr \tag{G1}$$

where we use the mode expansion (17) of the field $F$ and the fact that $F(0)=u'^2$ according to (7). Calculating the integral in the last expression above we have

$$L_{11} = \frac{2}{3u'^2}\sum_k E(k)\left\{\frac{3}{2k}\int_0^{k\pi}\frac{\sin x}{x}dx - \frac{\sin k\pi}{2k}\right\} \tag{G2}$$

The quantity in the brackets very quickly approaches the value $3\pi/4k$ (at $k\sim 5$ the difference is one per thousand), which gives the standard spectral formula for the integral scale, see e.g. [6]. Hence we see that the spectral formula for this quantity is somewhat different in compact space than the usual one. The reason for this fact is that, as opposed e.g. to the energy moments, which are given by local formulas in $r$-space, the integral scale is non-local, involving the integral of the field $F$. Also, note that given that $f(0)=1$ and $f(\pi)\sim 0$, the definition (G1) implies that $L_{11}$ should be certainly smaller than $\pi/2 \sim 1.6$. In practice where it turns out that $L_{11}\sim 1$ as we discussed in section VI, according to both the DNS and the results of the present work.